\documentclass[prc,showpacs,floatfix]{revtex4}
\usepackage{graphics}
\usepackage{psfrag}
\usepackage{times}
\usepackage{amsmath, amssymb}
\usepackage{subfigure,rotating}
\usepackage{bm}% bold math
\usepackage{bbold}
\usepackage{slashed}

\newcommand{\beq}{\begin{equation}}
\newcommand{\eeq}{\end{equation}}
\newcommand{\beqa}{\begin{eqnarray}}
\newcommand{\eeqa}{\end{eqnarray}}

\begin{document}
\title{
\hfill{\small {\bf MKPH-T-11-07}}\\
\bf Influence of Coulomb distortion on polarization observables in elastic electromagnetic
  hadron lepton scattering at low energies}
\author{Hartmuth Arenh\"ovel}
\affiliation{
Institut f\"ur Kernphysik,
Johannes Gutenberg-Universit\"at Mainz, D-55099 Mainz, Germany}
\date{\today}
\begin{abstract}
The formal expression for the most general polarization observable in
elastic electromagnetic lepton hadron scattering at low energies is
derived for the nonrelativistic regime. For the explicit evaluation
the influence of Coulomb distortion on various polarization
observables is calculated in a distorted wave Born
approximation. Besides the hyperfine interaction also the spin-orbit
interactions of lepton and hadron are included. For like charges the
Coulomb repulsion reduces strongly the size of polarization
observables compared to the plane wave Born approximation whereas for
opposite charges the Coulomb attraction leads to a substantial
increase of these observables for hadron lab kinetic energies below about
20 keV. 
\end{abstract}


\maketitle

%%%%%%%%%%%%%%%%%%%%%%%%%%%%%%%%%%%%%%%%%%%%%%%%%%%%% 
\section{Introduction}
%%%%%%%%%%%%%%%%%%%%%%%%%%%%%%%%%%%%%%%%%%%%%%%%%%%%%

Recently, Coulomb effects on polarization transfer from polarized
electrons or positrons to initially unpolarized protons or antiprotons
in elastic electromagnetic scattering have been studied in a distorted wave
approximation at low energies~\cite{Are07}. These studies were
motivated by the idea to polarize hadrons by their scattering on polarized
electrons or positrons in a storage ring~\cite{HoM94}. However, in
view of such a design it turned out that the considered observable,
i.e.\ the total cross section for the scattering of initially unpolarized
hadrons off polarized leptons to polarized final hadrons, the
polarization transfer $P_{z00z}$, cannot
contribute to a net polarization of the hadrons in a storage
ring. The reason for that is that this polarization observable does not 
contain a genuine hadronic spin-flip process~\cite{WaA07,MiS08}, which is
necessary for a net polarization change. Moreover, our previous numerical results were
critizised by Milstein et al.~\cite{MiS08} who had taken a partial
wave expansion of the Coulomb scattering wave function instead of the
integral representation used in ref.~\cite{Are07}. Indeed, it turned out that
besides a minor error the main reason for the gross overestimation of the
polarisation transfer cross section was an accuracy problem in the
numerical evaluation, namely,
the relevant quantity was calculated as a difference of two
almost equal numbers multiplied by a huge factor~\cite{Are08,WaA08}.  

For these reasons I have extended the previous study to the formal consideration of all
possible polarization observables in this scattering reaction including such
spin-flip transitions using again the distorted wave Born approximation.
In addition to the previously considered hyperfine interaction I have
included also the spin-orbit interactions of lepton and hadron. In the
next section the most general scattering cross section is introduced,
defining the various polarization observables in terms of bilinear
hermitean forms of the $T$-matrix elements. 
For the nonrelativistic form of the $T$-matrix with
inclusion of hyperfine and spin-orbit interactions the detailed
expression of the general scattering cross section is given, allowing for the
polarization of all initial and final particles described by
corresponding spin density matrices. 
In Section III I specialize to the case where the polarization of the
final lepton is not measured to the so-called triple polarization cross
section. For the numerical evaluation two different methods have been
applied, a partial wave expansion as in ref.~\cite{MiS08} and an
integral representation of the Coulomb wave function according to
ref.~\cite{LeA01}. Results for the structure functions and spin-flip
triple cross sections for the case of polarization along the incoming hadron
momentum are presented in Section IV and a summary
is given in Section V. Details for the evaluation of the hyperfine and
spin-orbit interactions are given in an appendix. 

%%%%%%%%%%%%%%%%%%%%%%%%%%%%%%%%%%%%%%%%%%%%%%%%%%%%%
\section{The general differential cross section including 
  polarization of  all particles} 
%%%%%%%%%%%%%%%%%%%%%%%%%%%%%%%%%%%%%%%%%%%%%%%%%%%%%

Reviews on polarization phenomena may be found for lepton hadron
scattering in~\cite{Dom69}, for nuclear physics in~\cite{Ohl72} and
for nucleon nucleon scattering in~\cite{ByL78}.
I will consider hadron-lepton scattering in
the c.m.\ system, where hadron stands for proton or antiproton and
lepton for electron or positron, 
\beq
h(\mathbf p \,) + l(-\mathbf p \,) \longrightarrow h(\mathbf
p^{\,\prime}) + l(-\mathbf p^{\,\prime}) \,, 
\eeq
allowing for inital and final hadron and lepton polarization. The
hadron initial and final three momenta are denoted by $\mathbf p$ and
$\mathbf p^{\,\prime}$, respectively. All possible observables of this
reaction can be obtained from the ``quadruple polarization'' cross
section for which the spin states of all initial and final particles
are described by the corresponding general spin density matrices
$\rho^{l/h}({\mathbf P}^{i/f}_{l/h})$, where the initial density matrices
characterize the spin properties of target and beam and the final ones
those of the detectors. It is given by the general trace 
\beqa
\frac{d\sigma^{quadruple}_{{\mathbf P}^{f}_{h},{\mathbf P}^{i}_{h},{\mathbf P}^{f}_{l},
 {\mathbf  P}^{i}_{l}}(\theta,\phi)}{d\Omega}&=&
{\cal O}({\mathbf P}^{f}_{h},{\mathbf P}^{f}_{l},{\mathbf P}^{i}_{h}
{\mathbf P}^{i}_{l};\theta,\phi) \nonumber\\
&=&\frac{M_l^2M_h^2}{\pi^2W^2(1+|{\mathbf P}^{f}_{l}|)(1+|{\mathbf P}^{f}_{h}|)}\,
\mbox{Trace}\Big[\widehat T^\dagger
\widehat \rho^{\, h}({\mathbf P}^{f}_{h})\widehat \rho^{\, l}({\mathbf P}^{f}_{l})\,
\widehat T\,\widehat \rho^{\, h}({\mathbf P}^{i}_{h})\widehat \rho^{\, l}({\mathbf P}^{i}_{l})\Big]\,,
\label{basic_xs}
\eeqa
where $\widehat T=\widehat T(\theta,\phi)$ denotes the T-matrix of the
scattering process with $(\theta,\phi)$ as scattering angles, and
$\rho(\mathbf P)$ the spin density matrix for a spin-$1/2$ particle,
with ${\mathbf P}$ characterizing the polarization of the 
corresponding particle in the initial and final states, respectively. 
The trace refers to the hadron and lepton spin degrees of freedom. The
factor in front takes into account the final phase space, the
incoming flux, and a normalization factor for the case of partially
polarized final states. The invariant energy of the hadron-lepton 
system is denoted by $W=E_h+E_l$ and the 
masses of hadron and lepton by $M_h$ and $M_l$, respectively.  
In the c.m.\ 
frame I use as reference system the $z$-axis along the incoming 
hadron momentum $\mathbf p$. The $x$- and $y$-axes are chosen to
form a right handed orthogonal system.  

In view
of the fact, that in this work I am interested in the low energy
regime, a nonrelativistic framework is adopted. 
The nonrelativistic density matrices for possible polarization of initial and
final states of a spin-$1/2$ particle have the standard form
\beqa
\widehat \rho\,({\mathbf P})&=&\frac{1}{2}(1+{\mathbf
  P}\cdot {\boldsymbol\sigma})\,.
\eeqa
with the vector ${\mathbf P}$ describing the polarization of the
particle and $\boldsymbol\sigma$ denoting the Pauli spin vector. One
should note that in general $|\mathbf P^{i/f}_{h/l}|\le 1$. 

From the basic equation
(\ref{basic_xs}) one obtains all possible polarization observables. 
In detail they are: 
\begin{description}
\item[(i)]
The unpolarized differential cross section:
\beq
\frac{d\sigma_0(\theta,\phi)}{d\Omega}={\cal O}
(\mathbf 0,\mathbf 0,\mathbf 0,\mathbf 0;\theta,\phi)
=S^0(\theta,\phi)\,.
\eeq
\item[(ii)]
Beam, target and beam-target asymmetries of the differential cross
section for unpolarized final states in the notation of
Bystricky et al.~\cite{ByL78}:
\beqa
\frac{d\sigma_{{\mathbf P}^{i}_{h},{\mathbf P}^{i}_{l}}(\theta,\phi)}{d\Omega}&=&
{\cal O}(\mathbf 0,\mathbf 0,{\mathbf P}^{i}_{h},{\mathbf
      P}^{i}_{l};\theta,\phi)\nonumber\\
&=&\frac{d\sigma_0(\theta,\phi)}{d\Omega}\Big(1+\sum_j
  P^{i}_{h,j}A_{00j0}(\theta,\phi) + \sum_k P^{i}_{l,k} A_{000k}(\theta,\phi)
+\sum_{j,k} P^{i}_{h,j} P^{i}_{l,k} A_{00jk} (\theta,\phi)\Big)\,,\label{beam-target-asy}
\eeqa
with the asymmetry vectors
\beqa
A_{00j0}(\theta,\phi)&=&\frac{1}{S^0}\frac{ \partial}{\partial {P}^{i}_{h,j}}
{\cal O}(\mathbf 0,\mathbf 0,{\mathbf P}^{i}_{h},\mathbf 0;\theta,\phi)\nonumber\\
&=&
\frac{1}{2S^0}\Big(
\frac{d\sigma_{{\mathbf P}^{i}_{h},{\mathbf P}^{i}_{l}}}{d\Omega}
-\frac{d\sigma_{-{\mathbf P}^{i}_{h},{\mathbf P}^{i}_{l}}}{d\Omega}
\Big)\Big|_{{P}^{i}_{h,k}=\delta_{jk}}\,,\\
A_{000j}(\theta,\phi)&=&\frac{1}{S^0}\frac{\partial}{\partial {P}^{i}_{l,j}}
{\cal O}(\mathbf 0,\mathbf 0,\mathbf 0,{\mathbf
      P}^{i}_{l};\theta,\phi)\nonumber\\
&=&\frac{1}{2S^0}\Big(
\frac{d\sigma_{{\mathbf P}^{i}_{h},{\mathbf P}^{i}_{l}}}{d\Omega}
-\frac{d\sigma_{{\mathbf P}^{i}_{h},-{\mathbf P}^{i}_{l,}}}{d\Omega}
\Big)\Big|_{{P}^{i}_{l,k}=\delta_{jk}}\,,
\eeqa
and the hadron-lepton asymmetry tensor 
\beqa
A_{00jk}(\theta,\phi)&=&\frac{1}{S^0}
\frac{\partial^2}{\partial {P}^{i}_{h,j}\partial {P}^{i}_{l,k}}
{\cal O}(\mathbf 0,\mathbf 0,{\mathbf P}^{i}_{h},{\mathbf
      P}^{i}_{l};\theta,\phi)\nonumber\\
&=&\frac{1}{4S^0}\Big(
\frac{d\sigma_{{\mathbf P}^{i}_{h},{\mathbf P}^{i}_{l}}}{d\Omega}
+\frac{d\sigma_{-{\mathbf P}^{i}_{h},-{\mathbf P}^{i}_{l}}}{d\Omega}
-\frac{d\sigma_{-{\mathbf P}^{i}_{h},{\mathbf P}^{i}_{l}}}{d\Omega}
-\frac{d\sigma_{{\mathbf P}^{i}_{h},-{\mathbf P}^{i}_{l}}}{d\Omega}\Big)
\Big|_{P^{i}_{h,m}=\delta_{jm},\, P^{i}_{l,n}=\delta_{kn}}\,.
\eeqa
\item[(iii)]
Polarization of the final lepton or hadron for unpolarized beam and target:
\beqa
P_{0j00}(\theta,\phi)&=&\frac{1}{S^0}\frac{\partial}{\partial {P}^{f}_{l,j}}
{\cal O}(\mathbf 0,{\mathbf P}^{f}_{l},\mathbf 0,\mathbf 0;\theta,\phi)\,,\\
P_{j000}(\theta,\phi)&=&\frac{1}{S^0}\frac{\partial}{\partial {P}^{f}_{h,j}}
{\cal O}({\mathbf P}^{f}_{h},\mathbf 0,\mathbf 0,\mathbf 0;\theta,\phi)\,.
\eeqa
\item[(iv)]
Various correlations between the polarization of one outgoing particle and beam
and/or target polarizations. For example, the outgoing hadron
polarization for initial lepton polarization but unpolarized incoming
hadron, the lepton-hadron polarization transfer is given by
\beqa
P_{j00k}(\theta,\phi)&=&\frac{1}{S^0}
\frac{\partial^2}{\partial {P}^{f}_{h,j}\partial {P}^{i}_{l,k}}
{\cal O}({\mathbf P}^{f}_{h},\mathbf 0 ,\mathbf 0,{\mathbf P}^{i}_{l};\theta,\phi)\,.
\eeqa
\item[(v)]
Another interesting example is the hadron spin-flip of an initially
polarized hadron by the scattering on an initially polarized lepton. It is
a special case of the so-called ``triple polarization'' cross section
with all particles polarized except for the final lepton as defined by
\beq
\frac{d\sigma^{triple}_{{\mathbf P}^{f}_{h},{\mathbf P}^{i}_{h},{\mathbf
      P}^{i}_{l}}(\theta,\phi)}{d\Omega}=
{\cal O}({\mathbf P}^{f}_{h},{\mathbf 0},{\mathbf P}^{i}_{h}
{\mathbf P}^{i}_{l};\theta,\phi)
\eeq
for the case ${\mathbf P}^{f}_{h}=-{\mathbf P}^{i}_{h}$,
i.e.\
\beq
\frac{d\sigma^{sf}_{{\mathbf P}^{i}_{h},{\mathbf P}^{i}_{l}}(\theta,\phi)}{d\Omega}=
\frac{d\sigma^{triple}_{-{\mathbf P}^{i}_{h},{\mathbf P}^{i}_{h},{\mathbf P}^{i}_{l}}(\theta,\phi)}{d\Omega}=
{\cal O}(-{\mathbf P}^{i}_{h},{\mathbf 0},{\mathbf P}^{i}_{h}
{\mathbf P}^{i}_{l};\theta,\phi)\,.
\eeq
This is the relevant quantity for the method of polarizing hadrons by
electromagnetic scattering on polarized leptons in a storage ring~\cite{WaA07,MiS08}.
\end{description}

\subsection{The nonrelativistic $T$-matrix}

For the explicit evaluation of the trace in eq.~(\ref{basic_xs}) one needs to know
the spin dependence of the $T$-matrix. In a nonrelativistic approach but
including contributions of the order $M^{-2}$, the
$T$-matrix contains the Coulomb, the lepton and hadron spin-orbit and
the lepton-hadron hyperfine interactions. Separating the various
contributions, the $T$-matrix is given in an obvious notation by 
\beqa
\widehat T&=&\widehat T_C+\widehat T_{LS_l}+\widehat T_{LS_h}+\widehat
T_{SS}\,.\label{tmatrix}  
\eeqa
In detail, one has the Coulomb contribution
\beqa
\widehat T_C=4\pi \alpha a_C\,,\label{tmatrix_C}
\eeqa
the spin-orbit interactions of lepton and hadron, respectively,
\beqa
\widehat T_{LS_{l/h}}&=&4\pi \alpha \mathbf b_{l/h}\cdot\boldsymbol\sigma^{l/h}\,,\label{tmatrix_so}
\eeqa
and the hyperfine interaction
\beqa
\widehat T_{SS}&=&4\pi \alpha 
\boldsymbol\sigma^h\cdot\stackrel{\leftrightarrow}{d}\cdot\boldsymbol\sigma^l\,,
\label{tmatrix_SS} 
\eeqa
where $\stackrel{\leftrightarrow}{d}$ denotes a symmetric rank-two tensor,  
$\mathbf q=\mathbf p^{\,\prime}-\mathbf p$ the three-momentum transfer, and
$\alpha$ denotes 
the Sommerfeld fine structure constant. The tensor
$\stackrel{\leftrightarrow}{d}$ can be decomposed 
into a scalar and a spherical tensor of rank two, i.e., a symmetric
cartesian tensor with vanishing trace,
\beq
{\stackrel{\leftrightarrow}{d}}=d^{[0]}+d^{[2]}\,,\label{d-separation}
\eeq
where
\beqa
d_{ij}^{[0]}&=&d_0\delta_{ij},\quad\mbox{ and}\quad d_0=\frac{1}{3}
\mbox{Trace}({\stackrel{\leftrightarrow}{d}})\,,\\
d_{ij}^{[2]}&=&d_{ij}-d_0\delta_{ij}\,.
\eeqa
Furthermore, the parameters $a_C$, $\mathbf b_{l/h}$, and
${\stackrel{\leftrightarrow}{d}}$ depend on what kind of approximation is used.
These are:
\begin{description}
\item[(i)]
Plane wave approximation (PW), corresponding to a pure one-photon exchange; the
nonrelativistic reduction of the $T$-matrix including lowest order
relativistic contribution reads
\beqa
\widehat T^{PW}&=&\frac{4\pi\alpha }{q^{2}}
\Big\{Z_lZ_h\Big(1 +\frac{{\mathbf P}^2}{4M_lM_h}\Big)
-\frac{1}{8} \Big( Z_h\frac{2\mu_l-1}{M_l^2}
+Z_l\frac{2\mu_h-1}{8M_h^2}\Big)q^{2}
\nonumber\\&&
-\frac{Z_h}{8M_l}\Big(\frac{2\mu_l-1}{M_l}+\frac{2\mu_l}{M_h}\Big) 
i(\boldsymbol\sigma_l\times\mathbf q\,)\cdot\mathbf P 
-\frac{Z_l}{8M_h}\Big(\frac{2\mu_h-1}{M_h}+\frac{2\mu_h}{M_l}\Big) 
i(\boldsymbol\sigma_h\times\mathbf q\,)\cdot\mathbf P
\nonumber\\ && 
+\frac{\mu_l\mu_h}{4M_lM_h}(\boldsymbol\sigma_l\cdot\mathbf q
\,\boldsymbol\sigma_h\cdot\mathbf q
-q^2\,\boldsymbol\sigma_l\cdot\boldsymbol\sigma_h
\,)\Big\}\,,
\eeqa
with $\mathbf P=\mathbf p+\mathbf p^{\,\prime}$, $Z_l$ and $Z_h$ as the lepton and hadron
charges, and $\mu_l$ and $\mu_h$ as their magnetic moments,
respectively.  From this expression one reads off the
parameters, keeping in the spin independent term the lowest order
only, 
\beqa
a_C^{PW}&=& \frac{Z_lZ_h}{q^2}\,,\\
\mathbf b^{\,PW}_{l/h}&=&i c_{l/h}^{LS}\,\frac{\mathbf p^{\,\prime}\times \mathbf p}{q^2}\,,\label{b-pw}\\
d_{ij}^{PW}&=&c^{SS}(\widehat q_i\widehat q_j -\delta_{ij})\,,
\eeqa
where $\widehat{q}$ denotes the unit vector along the three-momentum
transfer $\mathbf q$ and $q=|\mathbf q\,|$. The  separation  into a scalar and a traceless tensor according to
(\ref{d-separation}) reads
\beqa
d_{0}^{PW} &=& -\frac{2}{3}c^{SS}\,,\\
d_{ij}^{[2]\,PW}&=&c^{SS}(\widehat q_i\widehat q_j -\frac{1}{3}\delta_{ij})\,.
\eeqa
Furthermore, the strength parameters are
\beqa
c_l^{LS}&=&\frac{Z_h}{4M_l}\,\Big(\frac{2\mu_l-1}{M_l}+2\,
\frac{\mu_l}{M_h}\Big)\,,\\ 
c_h^{LS}&=&\frac{Z_l}{4M_h}\,\Big(\frac{2\mu_h-1}{M_h}+2\,
\frac{\mu_h}{M_l}\Big)\,,\\ 
c^{SS}&=&\frac{\mu_l \mu_h}{4M_l M_h}\,.
\eeqa
One should note that the strength parameter of the hadronic
spin-orbit interaction is about three orders of magnitude smaller than
the parameter of the leptonic one, because their ratio is
approximately
\beq
\frac{c_h^{LS}}{c_l^{LS}}\approx 2\mu_h\frac{M_l}{M_h}\approx 3\cdot 10^{-3}\,.
\eeq
\item[(ii)]
Distorted wave approximation (DW) using nonrelativistic Coulomb
scattering wave functions 
\beq
\psi^{C(+)}_{\mathbf p}(\mathbf
r\,)=\sqrt{\frac{\pi\eta_C}{\sinh{\pi\eta_C}}}
\,e^{-\frac{\pi}{2}\eta_C}\, e^{i\mathbf p\cdot\mathbf r}
\,_1F_1(-i\eta_C,1;i(pr-\mathbf p\cdot\mathbf r\,))\label{c-scatt-wave}
\eeq
and
\beq
\psi^{C(-)}_{\mathbf p}(\mathbf r\,)=\Big(\psi^{C(+)}_{-\mathbf p}(\mathbf r\,) \Big)^*\,,
\eeq
where 
$\psi^{C(\pm)}_{\mathbf p}$ denotes
the incoming and outgoing scattering waves~\cite{Mes69}, respectively. 
Here, $_1F_1(a,b;z)$ denotes the confluent hypergeometric
function.  In the expression for the scattering wave
in eq.~(\ref{c-scatt-wave}) I have already separated the constant Coulomb
phase factor $e^{i\sigma_C}$ with 
\beqa
\sigma_C&=&\arg[\Gamma(1+i\eta_C)]\,,\label{coulomb-phase}
\eeqa
because it will disappear in the observables. The relevant quantity
for Coulomb effects is the Sommerfeld Coulomb parameter
\beq
\eta_C=\alpha Z_l Z_h/v
\eeq 
with $v$ denoting the relative
hadron-lepton velocity.

Within this approach one finds
\beqa
a_C^{DW}&=& {e^{i\phi_C}}a_C^{PW}\,,\,\mbox{ with }\,
\phi_C(\theta)=-\eta_C \ln[\sin^2(\theta/2)]\,,\\
\mathbf b_{l/h}^{DW}&=&i\frac{c^{LS}_{l/h}}{4\pi}\int \frac{d^3r}{r^3}
 \psi^{C(-)}_{\mathbf p^{\,\prime}}(\mathbf r\,)^*
\,(\mathbf r\times \mathbf \nabla)\,
\psi^{C(+)}_{\mathbf p}(\mathbf r\,) \,,\label{ls_dwba}\\
d^{DW}_{ij}&=&-\frac{c^{SS}}{4\pi}\int d^3r
 \psi^{C(-)}_{\mathbf p^{\,\prime}}(\mathbf r\,)^*
\,\Big[\frac{1}{r^3}(3\hat{r}_i\,\hat{r}_j-\delta_{ij})+\frac{8\pi}{3}\delta_{ij}\delta(\mathbf
r\,) \Big]\,\psi^{C(+)}_{\mathbf  p}(\mathbf r\,)\,.
\label{t_dwba}
\eeqa
Separating again the hyperfine contribution into a scalar and a
traceless tensor,
one obtains
\beqa
d^{DW}_{0}&=&-\frac{2}{3}\,c^{SS}N(\eta_C)^2 \,,\\
d^{[2]\,DW}_{ij}&=&\frac{c^{SS}}{4\pi}\int \frac{d^3r}{r^3}
 \psi^{C(-)}_{\mathbf p^{\,\prime}}(\mathbf r\,)^*
\,(3\hat{r}_i\,\hat{r}_j-\delta_{ij})\,\psi^{C(+)}_{\mathbf  p}(\mathbf r\,)\,.
\label{t_dwba_2}
\eeqa
One should note that the tensor
$d_{ij}^{DW}$ in eq.~(\ref{t_dwba}) is symmetric as well as
$d^{[2]\,DW}_{ij}$.
\end{description}

\subsection{The general scattering cross section and polarization
  observables} 
Evaluation of the trace in eq.~(\ref{basic_xs}) yields the following general
expression 
\beqa\label{quadruplexs}
{\cal O}({\mathbf P}^{f}_{h},{\mathbf P}^{f}_{l},{\mathbf P}^{i}_{h}
{\mathbf P}^{i}_{l};\theta,\phi)
&=&\sum_{\alpha,\beta\in\{C,LS_l,LS_h,SS\}}
S_{\alpha,\beta}(\theta,\phi)\,, 
\eeqa
where the various contributions are defined by
\beq
S_{\alpha,\beta}(\theta,\phi)=\frac{M_l^2M_h^2}
{\pi^2W^2(1+|{\mathbf P}^{f}_{l}|)(1+|{\mathbf P}^{f}_{h}|)}
\mbox{Trace}\Big[\widehat T^\dagger_\alpha
\rho_h^f({\mathbf P}^{f}_{h})\rho_l^f({\mathbf P}^{f}_{l})
\widehat T_\beta\rho_h^i({\mathbf P}^{i}_{h})\rho_l^i({\mathbf P}^{i}_{l})\Big]\,
\eeq
with $\widehat T_\alpha$ defined in eqs.~(\ref{tmatrix_C})-(\ref{tmatrix_SS}). One
should note the relation 
\beq
S_{\alpha,\beta}=S_{\beta,\alpha}^*\,,
\eeq
from which follows that $S_\alpha:=S_{\alpha,\alpha}$ is
real. 

Separating the diagonal contributions ($S_\alpha$) from the
interference terms ($S_{\alpha,\beta}$ for $\alpha\neq\beta$), one
obtains  for the ``quadruple polarization'' cross section
\beqa
\frac{d\sigma_{{\mathbf P}^{f}_{h},{\mathbf P}^{f}_{l},{\mathbf P}^{i}_{h},
{\mathbf P}^{i}_{l}}(\theta,\phi)}{d\Omega}
&=&\sum_{\alpha\in\{C,LS_l,LS_h,SS\}}
S_{\alpha}(\theta,\phi)+
\sum_{\alpha<\beta\in\{C,LS_l,LS_h,SS\}}2\,\,Re\,
S_{\alpha,\beta} (\theta,\phi)\,.
\eeqa
Explicitly, one finds in terms of the different contributions to the
$T$-matrix in eq.~(\ref{tmatrix}) for the diagonal terms
\beqa
S_{C}(\theta,\phi)&=&V_0\,|a_C|^2\Pi_h^+\Pi_l^+\,,\label{S_C}\\
S_{LS_l}(\theta,\phi)&=&V_0\,\Pi_h^+\Big(\Pi_l^-\,\mathbf b^*_{l}\cdot \mathbf b_{l}
+2\,Re \Big[(\mathbf b_{l}\cdot\mathbf P^f_l)^*
(\mathbf b_{l}\cdot \mathbf P^i_l) \Big]\Big)\,,\label{S_LSl}\\
S_{LS_h}(\theta,\phi)&=&S_{LS_l}(\theta,\phi) |_{h\leftrightarrow l}\,,\label{S_LSh}\\
S_{SS}(\theta,\phi)&=&V_0\,\Big(\Pi_h^-\Pi_l^-D_0
-\mathbf P_{h}^-\cdot\stackrel{\leftrightarrow}{G} \cdot \mathbf P_{l}^-
+\Big[ \Pi_h^-\mathbf P^f_l\cdot{\stackrel{\leftrightarrow}{D}} \cdot\mathbf
P^i_l +(l\leftrightarrow h) \Big]\nonumber\\&&
-\,Im \Big[ \Big(\Pi_h^- (\mathbf P^{-}_l\cdot \mathbf H) 
+2\sum_{jst} P_{l,j}^- P_{h,s}^{f} P_{h,t}^{i}\,E_{jst}\Big)
+(l\leftrightarrow h) \Big]\nonumber\\
&&
+2\,Re \Big[(\mathbf P_{l}^f\cdot{\stackrel{\leftrightarrow}{d}}^*\cdot\mathbf P^f_{h})\,
(\mathbf P_{l}^i\cdot{\stackrel{\leftrightarrow}{d}}\cdot\mathbf P^i_{h})
+(\mathbf P_{l}^i\cdot{\stackrel{\leftrightarrow}{d}}^* \cdot\mathbf P^f_{h})\,
(\mathbf P_{h}^i\cdot{\stackrel{\leftrightarrow}{d}}\cdot\mathbf P^f_{l}) \Big]
\Big)\,,\label{S_SS}
\eeqa
where 
\beq
V_0=\frac{4\alpha^2M_l^2M_h^2}{W^2(1+|{\mathbf P}^{f}_{l}|)(1+|{\mathbf P}^{f}_{h}|)}\,. \label{vorfactor}
\eeq
The following quantities depend on the polarization parameters
\beqa
\Pi_{h/l}^{\pm}&=&1\pm\mathbf P^f_{h/l}\cdot \mathbf P^i_{h/l}\,,\label{pol-op1}\\
\mathbf P^{\pm}_{h/l}&=&\mathbf P^i_{h/l}\pm \mathbf P^f_{h/l}\,,\\
{\mathbf Q}_{h/l}&=&\mathbf P^{+}_{h/l}-i\mathbf P^f_{h/l}\times \mathbf P^i_{h/l}\label{pol-op3}\,.
\eeqa
Furthermore, I have introduced  for convenience the following
quantities, which depend on the hyperfine interaction tensor $d_{ij}$,
\beqa
D_{ij}&=&2\,Re\Big(\sum_k d_{ik}^*
d_{kj}\Big)\,,\\ 
D_0&=&\sum_{ij} d_{ij}^*d_{ji}=
\frac{1}{2}\mbox{Trace}(\stackrel{\leftrightarrow}{D})\,,\\
E_{jst}&=&\sum_{kl}\varepsilon_{jkl}\, d_{ks}^*d_{lt}\,,\\ 
G_{ij}&=&
\sum_{lmst}\varepsilon_{ils}\,\varepsilon_{jmt}\,d_{lm}^*d_{st}\,,\\ 
H_{i}&=&
\sum_{klm}\varepsilon_{ikl}\,d_{km}^*d_{ml}\,,
\eeqa
where $\varepsilon_{ikl}$ denotes the totally antisymmetric
Levi-Civita tensor in three dimensions. 
These quantities are functions which depend on the scattering angles
$(\theta,\phi)$. 

 Correspondingly, one finds for the interference terms
\beqa 
S_{C,LS_l}(\theta,\phi)&=&V_0\,a_C^* \Pi_h^+\,\mathbf b_{l}\cdot\mathbf Q_l\,,\\
S_{C,LS_h}(\theta,\phi)&=&S_{C,LS_l}(\theta,\phi)|_{h\leftrightarrow l}\,,\\
S_{C,SS}(\theta,\phi)&=&V_0\,a_C^*\,\mathbf Q_h\cdot{\stackrel{\leftrightarrow}{d}} \cdot \mathbf Q_l\,,\\
S_{LS_l,LS_h}(\theta,\phi)&=&V_0\,(\mathbf b_{l}\cdot\mathbf Q_l) ^*\,(\mathbf b_{h}\cdot\mathbf Q_h)\,,\\
S_{LS_l,SS}(\theta,\phi)&=&V_0\,\mathbf Q_h\cdot{\stackrel{\leftrightarrow}{d}} \cdot\Big(\Pi_l^-\,\mathbf b_l^* -i\,\mathbf
b_l^*\times\mathbf P^{-}_l %\nonumber\\&&
+ (\mathbf b_{l}^*\cdot\mathbf P^i_l)\,\mathbf P^f_l+(\mathbf b_{l}^*\cdot\mathbf P^f_l)\,\mathbf P^i_l\Big)\,.\label{S_LS_l_SS}\\
S_{LS_h,SS}(\theta,\phi)&=&S_{LS_l,SS}(\theta,\phi)|_{h\leftrightarrow l}\,.
\eeqa

Using the separation of the tensor
$\stackrel{\leftrightarrow}{d}$ 
into a scalar and a traceless symmetric tensor according to
eq.~(\ref{d-separation}), one finds
\beqa
D_0&=&3|d_0|^2+\sum_ {i,k} d^{[2]*}_{ik} d^{[2]}_{ki}\,,\\ 
D_{ij}&=&2|d_0|^2\delta_{ij}+4\,Re(d_0^*d^{[2]}_{ij})+D_{ij}^{(2)}\,,\\ 
D_{ij}^{(2)} &=&2\,Re \sum_ k d^{[2]*}_{ik} d^{[2]}_{kj}
\,,\\ 
E_{jst}&=&\varepsilon_{jst}\,|d_0|^2+\sum_{kl}\varepsilon_{jkl}\,d_{ks}^{[2]\,*}d_{lt}^{[2]}
+\Big(\sum_{l}\varepsilon_{jsl}\,d_0^*
d^{[2]}_{lt}-(s\leftrightarrow t)^*\Big)
\,,\\  
G_{ij}&=&2|d_0|^2\delta_{ij}-2\,Re
(d_0^*d^{[2]}_{ij})+G_{ij}^{(2)}\,,\\ 
H_{i}&=&
\sum_{klm}\varepsilon_{ikl}\,d^{[2]*}_{km} d^{[2]}_{ml}\,,\eeqa
where
\beqa
G_{ij}^{(2)}&=&
\sum_{lmst}\varepsilon_{ils}\varepsilon_{jmt}d_{lm}^{[2]*}d_{st}^{[2]}\,. 
\eeqa
It is now easy to see that the vector $\mathbf H$ is purely imaginary and
that the tensor $G_{ij}$ is real and symmetric. Furthermore, one notes the
symmetry property
\beq
E_{jst}^*=-E_{jts}\,.
\eeq
It suffices to evaluate the spin-orbit vector $\mathbf b$ and the hyperfine
tensor $\stackrel{\leftrightarrow}{d}$ for $\phi=0$, because
then the values for an arbitrary $\phi$ can be generated by a rotation
around the $z$-axis exploiting their rotation properties.  Examples
are given in the following section.

\section{The triple polarization cross section}

I will now specialize to the case where only the final lepton polarization
is not analyzed, i.e.\ $\mathbf P^f_l=0$, but all other particles are
completely polarized ($|\mathbf P^{i/f}_h|=1,\, |\mathbf P^{i}_l|=1$).  
This case is of particular
interest for the polarization transfer in a storage ring~\cite{WaA07,MiS08}. The
corresponding ``triple polarization'' cross section has the form
\beqa\label{triplexs}
\frac{d\sigma^{triple}_{{\mathbf P}^{f}_{h},{\mathbf P}^{i}_{h}
{\mathbf P}^{i}_{l}}(\theta,\phi)}{d\Omega}&=&
{\cal O}({\mathbf P}^{f}_{h},{\mathbf 0},{\mathbf P}^{i}_{h}
{\mathbf P}^{i}_{l};\theta,\phi)\nonumber\\
&=&
\sum_{\alpha\in\{C,LS_l,LS_h,SS\}}
S_{\alpha}^{triple} (\theta,\phi)+
\sum_{\alpha<\beta\in\{C,LS_l,LS_h,SS\}}
2\,Re \,S_{\alpha,\beta}^{triple} (\theta,\phi)\,.
\eeqa
In this case, the lepton polarization
quantities in (\ref{pol-op1}) through (\ref{pol-op3}) become
\beq
\Pi_l^{\pm}=1,\quad \mathbf Q_l=\mathbf P_l^i,\quad \mathbf P_l^{\pm}=\mathbf P_l^i\,,
\eeq
and one finds for the diagonal terms
\beqa
S_C ^{triple}(\theta)&=&V\,\Pi_h^+|a_C (\theta)|^2\,,\\
S_{LS_l}^{triple} (\theta,\phi)&=&V\,\Pi_h^+\,\mathbf b^*_{l}\cdot \mathbf b_{l}\,,\\
S_{LS_h}^{triple} (\theta,\phi)&=&V\,\Big(\Pi_h^-\,\mathbf b^*_{h}\cdot \mathbf b_{h}
+2\,Re \Big[(\mathbf b_{h}\cdot\mathbf P^f_h)^*
(\mathbf b_{h}\cdot \mathbf P^i_h) \Big]\Big)\,,\\
S_{SS}^{triple} (\theta,\phi)&=&V\,\Big(\Pi_h^-D_0
-\mathbf P_{h}^-\cdot\stackrel{\leftrightarrow}{G} \cdot \mathbf P_{l}^i +\mathbf P^f_h\cdot{\stackrel{\leftrightarrow}{D}} \cdot\mathbf P^i_h 
\nonumber\\&&
-\,Im \Big[\mathbf P^{-}_h\cdot \mathbf H +\Pi_h^-(\mathbf P^{i}_l\cdot \mathbf H)
+2\sum_{jst} P_{l,j}^i P_{h,s}^{f} P_{h,t}^{i}\,E_{jst}
\Big]\Big)\,,
\eeqa
and for the interference terms
\beqa
S_{C,LS_l}^{triple} (\theta,\phi)&=&V\,a_C^*\Pi_h^+\,\mathbf b_{l}\cdot\mathbf P_l^i\,,\\
S_{C,LS_h}^{triple} (\theta,\phi)&=&V\,a_C^*\,\mathbf b_{h}\cdot\mathbf Q_h\,,\\
S_{C,SS}^{triple}(\theta,\phi)&=&V\,a_C^*\,\mathbf Q_h\cdot{\stackrel{\leftrightarrow}{d}} \cdot \mathbf P_l^i\,,\\
S_{LS_l,LS_h}^{triple} (\theta,\phi)&=&V\,(\mathbf b_{l}\cdot\mathbf P_l^i) ^*\,(\mathbf b_{h}\cdot\mathbf Q_h)\,,\\
S_{LS_l,SS}^{triple} (\theta,\phi)&=&V\,\mathbf Q_h\cdot{\stackrel{\leftrightarrow}{d}} \cdot\Big(\mathbf b_l^* -i\,\mathbf
b_l^*\times\mathbf P^{i}_l \Big) \,,\\
S_{LS_h,SS}^{triple} (\theta,\phi)&=&V\,\mathbf
P_l^i\cdot{\stackrel{\leftrightarrow}{d}} \cdot\Big(\Pi_h^-\,\mathbf b_h^*
-i\,\mathbf b_h^*\times\mathbf P^{-}_h 
+ (\mathbf b_{h}^*\cdot\mathbf P^i_h)\,\mathbf P^f_h+(\mathbf b_{h}^*\cdot\mathbf P^f_h)\,\mathbf P^i_h\Big)\,.
\eeqa
where 
\beq
V=\frac{2\alpha^2M_l^2M_h^2}{W^2}\,. \label{vorfactor_a}
\eeq
 From now on as a further specialization, I will consider
only polarization along the incoming direction which is chosen as
$z$-axis. Then with
\beq
\mathbf P_h^{i/f}=\lambda_h^{i/f}\hat z\,,\quad \mathbf P_h^{\pm}=
(\lambda_h^i\pm\lambda_h^f) \hat z\,,\,\,\,\mbox{ and }\,\,\, \Pi_h^{\pm}=1\pm 
\lambda_h^i\lambda_h^f\,,
\eeq
where $\lambda_h^\pm=\lambda_h^i\pm\lambda_h^f$, one obtains 
\beqa
\frac{d\sigma ^{triple}_{\lambda^f_h,\lambda^i_h,\lambda^i_l}(\theta,\phi)}{d\Omega}&=&
(1+\lambda_h^i\lambda_h^f)\Big[S_C(\theta)+S_0(\theta,\phi)+L_0^l(\theta,\phi)\Big]
+(1-\lambda_h^i\lambda_h^f)\Big[L_0^h(\theta,\phi)+\lambda_l^i \Big (S_1(\theta,\phi)+L_1^{h}(\theta,\phi) \Big)\Big]\nonumber\\
&&+\lambda_h^-\Big[S_1(\theta,\phi)+\lambda_l^i \Big (S_2^-(\theta,\phi)+L_2^{h}(\theta,\phi) \Big)\Big]
+\lambda_h^+\Big[\lambda_l^i \Big (S_{2}^+(\theta,\phi) +L_2^{l}(\theta,\phi)\Big)+L_1^{l}(\theta,\phi\Big] \nonumber\\&&
+\lambda_h^i\lambda_h^f\Big[ S_2(\theta,\phi)+\lambda_l^iS_3(\theta,\phi) \Big]
\,.\label{hadron}
\eeqa
The diagonal
contributions are
\beqa
L_0^{l/h}(\theta,\phi)&=&V|\mathbf b_{l/h} (\theta,\phi)|^2 \,,\label{L0_lh}\\
S_C(\theta)&=&V|a_C|^2\,,\\
S_0(\theta,\phi)&=&V\Big(3|d_0|^2+\sum_j D_{jj}(\theta,\phi) \Big)\,,\\
S_1(\theta,\phi)&=&iV\sum_{jk}\epsilon_{3jk}D_{jk}(\theta,\phi)
=i(D_{12}(\theta,\phi)-D_{21}(\theta,\phi))=-2\,Im(D_{12}(\theta,\phi))\,,\\
S_2^-(\theta,\phi)&=&2V\Big(\,Re\Big(d_0^*d^{[2]}_{33}(\theta,\phi)
-d^{[2]}_{11}(\theta,\phi)^*
d^{[2]}_{22}(\theta,\phi)\Big) -|d_0|^2+|d^{[2]}_{12}(\theta,\phi)|^2\Big)\,,\\
S_2(\theta,\phi)&=&2V\Big(2\,Re
\Big(d_0^*d^{[2]}_{33}(\theta,\phi)\Big) 
-2|d_0|^2-D_{11}(\theta,\phi)-D_{22}(\theta,\phi)\Big)\,,\\
S_3(\theta,\phi)&=&-4V\,Im
\Big(d^{[2]}_{31}(\theta,\phi)^*d^{[2]}_{32}(\theta,\phi)\Big)\,,\label{S3} 
\eeqa
and the interference terms
\beqa
L_1^{l/h}(\theta,\phi)&=&2V\,Re\Big(d^{[2]}_{31}(\theta,\phi)\,b^*_{l/h,1}
(\theta,\phi)+d^{[2]}_{32}(\theta,\phi)\,b^*_{l/h,2} (\theta,\phi)\Big) \,,\\
L_2^{l/h}(\theta,\phi)&=&2V\,Im\Big(d^{[2]}_{31}(\theta,\phi)\,b^*_{l/h,2}
(\theta,\phi)-d^{[2]}_{32}(\theta,\phi)\,b^*_{l/h,1} (\theta,\phi)\Big) \,,\\
S_2^+(\theta,\phi)&=&2V\,Re\Big[a_C^*(d_0+d^{[2]}_{33}(\theta,\phi))\Big]\,.\label{S2plus}
\eeqa
Here I have already used the fact that the spin-orbit vector $\mathbf
b_{l/h}$ is perpendicular to the $z$-axis, i.e.\ its third
component vanishes. Indeed, as shown explicitly in the appendix, it
has the form
\beq
\mathbf b_{l/h} (\theta,\phi)=i\,b_0^{l/h}(\theta)(-\sin{\phi},
\cos{\phi},0)=
i\,b^{l/h}_0 (\theta)\,\frac{\mathbf p^{\,\prime}\times\mathbf p}{|\mathbf p^{\,\prime}\times\mathbf p\,|}\,.\label{bls}
\eeq

However, not all of the contributions in eqs.~(\ref{L0_lh}) through
(\ref{S2plus}) are nonzero. In 
fact, I now will show that the diagonal contributions 
$S_1$ and $S_3$ and the interference terms  $L_1^{l/h}$ vanish
identically. To this end I first will
consider the $\phi$-dependence of the tensors $d^{[2]}(\theta,\phi)$ and
$D(\theta,\phi)$. It suffices to evaluate them for $\phi=0$ from which one
obtains $ d^{[2]} (\theta,\phi)$ and $D(\theta,\phi)$ by a rotation around the
$z$-axis by an angle $\phi$. Introducing the notation
$d^{[2],0}(\theta)= d^{[2]} (\theta,\phi=0)$ and correspondingly
$D^0= D(\theta,0)$, and using furthermore the
fact $ d^{[2],0}_{12}= d^{[2],0}_{21}=0$,  one finds
\beq
d^{[2]} (\theta,\phi)=\left(
\begin{matrix}
d^{[2],0}_{11}(\theta)\cos^2\phi+d^{[2],0}_{22}(\theta)\sin^2\phi & 
\frac{1}{2}(d^{[2],0}_{11}(\theta)-d^{[2],0}_{22}(\theta))\sin 2\phi & 
d^{[2],0}_{13}(\theta)\cos\phi\cr 
& & \cr
\frac{1}{2}(d^{[2],0}_{11}(\theta)-d^{[2],0}_{22}(\theta))\sin 2\phi & 
d^{[2],0}_{11}(\theta)\sin^2\phi
+d^{[2],0}_{22}(\theta)\cos^2\phi & d^{[2],0}_{13}(\theta)\sin\phi\cr
& & \cr
 d^{[2],0}_{13}(\theta)\cos\phi & d^{[2],0}_{13}(\theta)\sin\phi & d^{[2],0}_{33}(\theta)\cr
\end{matrix}\right).\label{d2phi}
\eeq
From this representation one notes immediately that
\beq
d^{[2]}_{31}(\theta,\phi)^* d^{[2]}_{32}(\theta,\phi)= |d^{[2],0}_{13}(\theta)|\cos\phi\sin\phi
\eeq
is real and thus $S_3(\theta,\phi)=0$ according to eq.~(\ref{S3}). 
Furthermore, with
\beq
D^0(\theta)=\left(
\begin{matrix}
|d^{[2],0}_{11}(\theta)|^2+|d^{[2],0}_{13}(\theta)|^2 & 0 & 
d^{[2],0}_{11}(\theta)^*d^{[2],0}_{13}(\theta)+d^{[2],0}_{13}(\theta)^*d^{[2],0}_{33}(\theta)\cr
& & \cr
0 & |d^{[2],0}_{22}(\theta)| & 0 \cr
& & \cr
d^{[2],0}_{13}(\theta)^*d^{[2],0}_{11}(\theta)+d^{[2],0}_{33}(\theta)^*d^{[2],0}_{13}(\theta) & 
0 & |d^{[2],0}_{33}(\theta)|^2+|d^{[2],0}_{13}(\theta)|^2\cr
\end{matrix}\right),\label{E0}
\eeq
which formally has the same structure as $d^{[2],0}(\theta)$, one finds 
\beq
D(\theta,\phi)=\left(
\begin{matrix}
D^0_{11}(\theta)\cos^2\phi+D^0_{22}(\theta)\sin^2\phi & 
\frac{1}{2}(D^0_{11}(\theta)-D^0_{22}(\theta))\sin
2\phi & D^0_{13}(\theta)\cos\phi\cr
& & \cr
\frac{1}{2}(D^0_{11}(\theta)-D^0_{22}(\theta))\sin 2\phi & 
D^0_{11}(\theta)\sin^2\phi
+D^0_{22}(\theta)\cos^2\phi & D^0_{13}(\theta)\sin\phi\cr
& & \cr
 D^0_{13}(\theta)\cos\phi & D^0_{13}(\theta)\sin\phi & D^0_{33}(\theta)\cr
\end{matrix}\right).\label{D0}
\eeq
Here again one notes that
\beq
D_{12}(\theta,\phi)= \frac{1}{2}(D^0_{11}(\theta)-D^0_{22}(\theta))\sin 2\phi 
\eeq
is real because according to eq.~(\ref{E0}) $D^0_{11}(\theta)$ and
$D^0_{22}(\theta)$ are real and, therefore, $S_1=0$. 
For the spin-orbit interaction vector $\mathbf b_{l/h}$ one finds from eq.~(\ref{bls})
$|\mathbf b_{l/h} (\theta,\phi)|^2=|b_0^{l/h}(\theta)|^2$, which means
that $L_0^{l/h}$ is $\phi$-independent. Moreover, using
eqs.~(\ref{bls}) and (\ref{d2phi}) one obtains 
\beq
L_1^{l/h}(\theta,\phi)=4V\,Re\Big(d^{[2],0}_{13}(\theta)\cos\phi (-b_0^{l/h}(\theta)\sin{\phi})   
+d^{[2],0}_{13}(\theta)\sin\phi \,b_0^{l/h}(\theta)\cos{\phi}\Big)=0\,.
\eeq

Finally, with the help of eqs.~(\ref{d2phi}) through (\ref{D0}) one finds that
the remaining structure functions become $\phi$-independent, which is
easy to understand since all polarizations are assumed to be along the
$z$-axis ruling out any $\phi$-dependence. Thus 
also the cross section simplifies to 
\beqa
\frac{d\sigma ^{triple}_{\lambda^f_h,\lambda^i_h,\lambda^i_l}(\theta,\phi)}{d\Omega}&=&
(1+\lambda_h^i\lambda_h^f)\Big[S_C(\theta)+S_0(\theta)+L_0^{l} (\theta) \Big]
+(1-\lambda_h^i\lambda_h^f)L_0^{h} (\theta)\nonumber\\
&&+\lambda_l^i\Big[ \lambda_h^- \Big(S_2^-(\theta,\phi)+L_2^{h}(\theta) \Big)
+\lambda_h^+\Big (S_{2}^+(\theta)  +L_2^{l}(\theta) \Big)\Big]%\nonumber\\&&
+\lambda_h^i\lambda_h^fS_2(\theta)\,,\label{hadron1}
\eeqa
where the structure functions are given by
\beqa
L_0^{l/h}(\theta)&=&V|b_0^{l/h} (\theta)|^2 \,,\\
L_2^{l/h}(\theta) &=& 2V\,Re\Big[d^{[2],0}_{13}(\theta) b_0^{l/h}(\theta)^* \Big] \,,\\
S_0(\theta)&=&V \Big (3|d_0|^2+\sum_{j=1}^3|d^{[2],0}_{jj}(\theta)|^2
+2|d^{[2],0}_{13}(\theta)|^2 \Big)\,,\\
S_2^+(\theta)&=&2V\,Re \Big [a_C^*\Big (d_0+d^{[2],0}_{33}(\theta) \Big) \Big]\,,\\
S_2^-(\theta)&=&2V\Big[\,Re \Big (d_0^*d^{[2],0}_{33}(\theta)-d^{[2],0}_{11}(\theta)^*
d^{[2],0}_{22}(\theta) \Big)-|d_0|^2\Big]\,,\\
S_2(\theta)&=&2V\Big[2\,Re \Big (d_0^*d^{[2],0}_{33}(\theta) \Big) 
-2|d_0|^2-|d^{[2],0}_{11}(\theta)|^2
-|d^{[2],0}_{22}(\theta)|^2-|d^{[2],0}_{13}(\theta)|^2
\Big]\,.
\eeqa
The expression in eq.~(\ref{hadron1}) is an extension of the triple
polarization cross section given
in eq.~(16) of ref.~\cite{WaA07} by including the contributions from
the hadron and lepton spin-orbit interactions. (As a sideremark: The
corresponding cross section for final lepton polarization is obtained
from (\ref{hadron}) by the substitutions $\lambda_h^f\to\lambda_l^f$ and
$\lambda_h^i\leftrightarrow\lambda_l^i$.)

As a special case, one considers the 
so-called hadronic spin-flip cross section for complete hadron
polarization, i.e.\
$\lambda_h^i=-\lambda_h^f=\lambda_h=\pm 1$, and the non-spin-flip
cross section with $\lambda_h^i=\lambda_h^f=\lambda_h=\pm 1$. 
For the spin-flip cross section one finds
\beqa
\frac{d\sigma^\mathrm{sf}_{-\lambda_h,\lambda_h,\lambda^i_l}(\theta,\phi)}{d\Omega}&=&
2 L_0^{h} (\theta) -S_2(\theta) 
+2\lambda_l^i\lambda_h\Big[S_2^-(\theta,\phi)+L_2^{h}(\theta) \Big]
\,.\label{hadron-flip}
\eeqa
It is governed by the  hyperfine terms $S_2$ and $S_2^-$ and
the hadronic spin-orbit interaction via $L^h_0$ and $L^h_2$. On the other hand, 
the non-spin-flip cross section is given by
\beqa
\frac{d\sigma^\mathrm{nsf}_{\lambda_h,\lambda_h,\lambda^i_l}(\theta,\phi)}{d\Omega}&=&
2 \Big[S_C(\theta)+S_0(\theta)+L_0^{l} (\theta) \Big] +S_2(\theta) 
+2\lambda_l^i\lambda_h\Big[S_2^+(\theta,\phi)+L_2^{l}(\theta) \Big]
\,.\label{hadron-nonflip}
\eeqa
Its helicity independent part is overwhelmingly dominated by the
Coulomb term $S_C$ with additional tiny contributions 
from the hyperfine and the leptonic spin-orbit interactions. 
The difference of the non-spin-flip cross section for $\lambda_h=\pm 1$
\beqa
\frac{1}{2}\,\Big(\frac{d\sigma^\mathrm{nsf}_{1,1,\lambda}(\theta,\phi)}{d\Omega} 
-\frac{d\sigma^\mathrm{nsf}_{-1,-1,\lambda}(\theta,\phi)}{d\Omega} \Big)
&=&2 \lambda\Big[S_2^+(\theta,\phi)+L_2^{l}(\theta) \Big]\,\label{diff-nonflip}
\eeqa
has been considered  as lepton-hadron polarization transfer in~\cite{HoM94}. 
This polarization transfer is dominated by the hyperfine structure
function $S_2^+$ because the additional spin-orbit contribution
$L_2^l$ is comparably small as shown in the next section.  It  differs
by a factor 2 and the presence of  $L_2^l$ from the polarization
transfer $P_{z00z}$ for the scattering of unpolarized hadrons on
polarized leptons as considered in ref.~\cite{Are07}, where I had
considered only the leading term $S_2^+$, 
the interference between Coulomb and hyperfine amplitudes, neglecting
higher order contributions.  The more complete expression reads
\beqa
P_{z00z}\frac{d\sigma_0(\theta,\phi)}{d\Omega}&=&
\frac{\partial^2}{\partial\lambda^f_h \partial\lambda^i_l}
\frac{d\sigma^{triple}_{\lambda^f_h,\lambda^i_h,\lambda^i_l}(\theta,\phi)}{d\Omega} 
\Big|_{\lambda^i_h=0}\nonumber\\
&=&S_2^+(\theta,\phi)-S_2^-(\theta,\phi)+L_2^l(\theta,\phi)-L_2^h(\theta,\phi)\,.
\eeqa
In addition to  $S_2^+$, it includes $S_2^-$, which is quadratic
in the hyperfine amplitude, and $L_2^l$ and $L_2^h$, the contributions from the
interference of hyperfine and leptonic
and hadronic spin-orbit amplitudes,
respectively. However, the largest of these additional terms, $L_2^l$,
is still quite small if not negligible compared to $S_2^+$.

\section{Results for structure functions and polarization cross sections}

For the evaluation of the structure functions in eq.~(\ref{hadron1})
and the corresponding cross section I have used both
methods for the calculation of Coulomb distortion, the integral
representation as well as the partial 
wave expansion. The integral representation has been used mainly in
order to check the convergence of the partial wave expansion as is
described in detail in the appendix. Thus all results presented in
this section are based on the partial wave expansion (PWE). For the
hyperfine amplitude it was found that an expansion up to a partial
wave with $l_{max}=2000$ was sufficient, but for the spin-orbit interaction,
beeing much slower convergent, $l_{max}=4000$ was taken.

\subsection{The structure functions}

\begin{figure}
\includegraphics[width=1.\columnwidth]{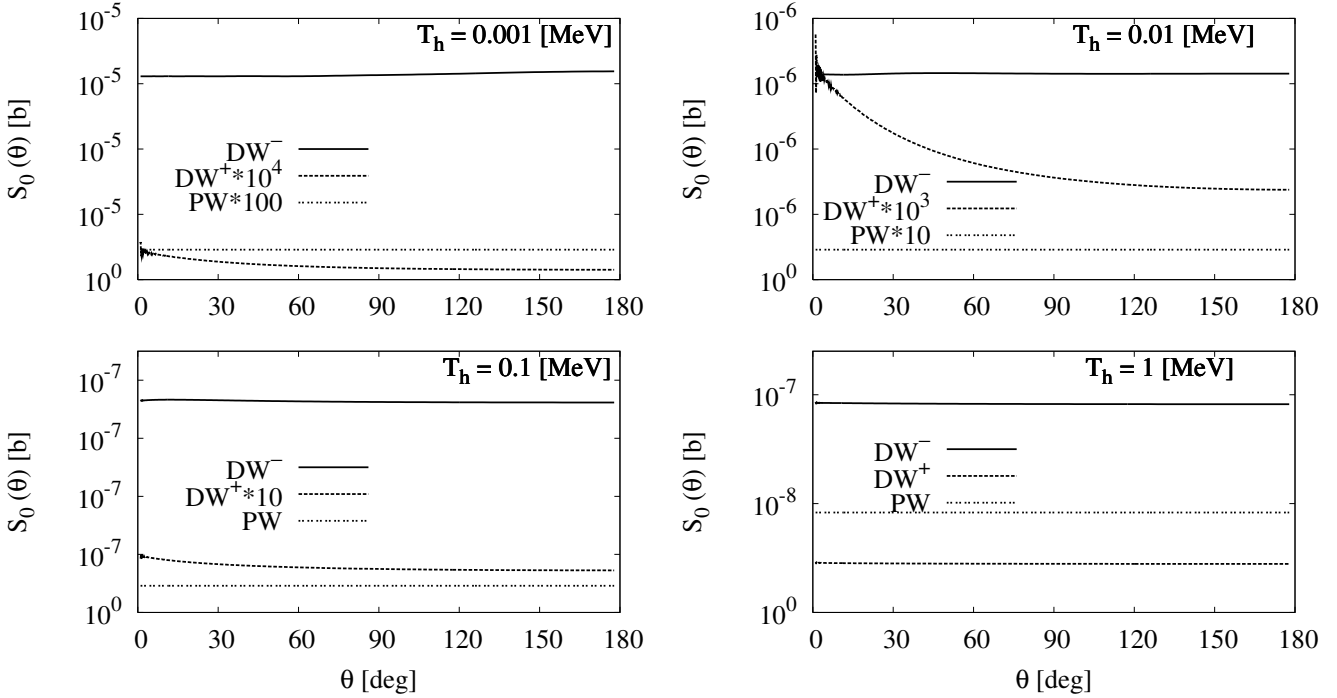}
\caption{Structure function $S_0(\theta)$ of the triple polarization
  differential cross section
  in plane wave (PW) and distorted wave approximation
  for like charges (DW$^+$) and opposite charges (DW$^-$) 
 for various proton lab kinetic energies $T_h$. }
\label{fig_S_0} 
\end{figure}

\begin{figure}
\includegraphics[width=1.\columnwidth]{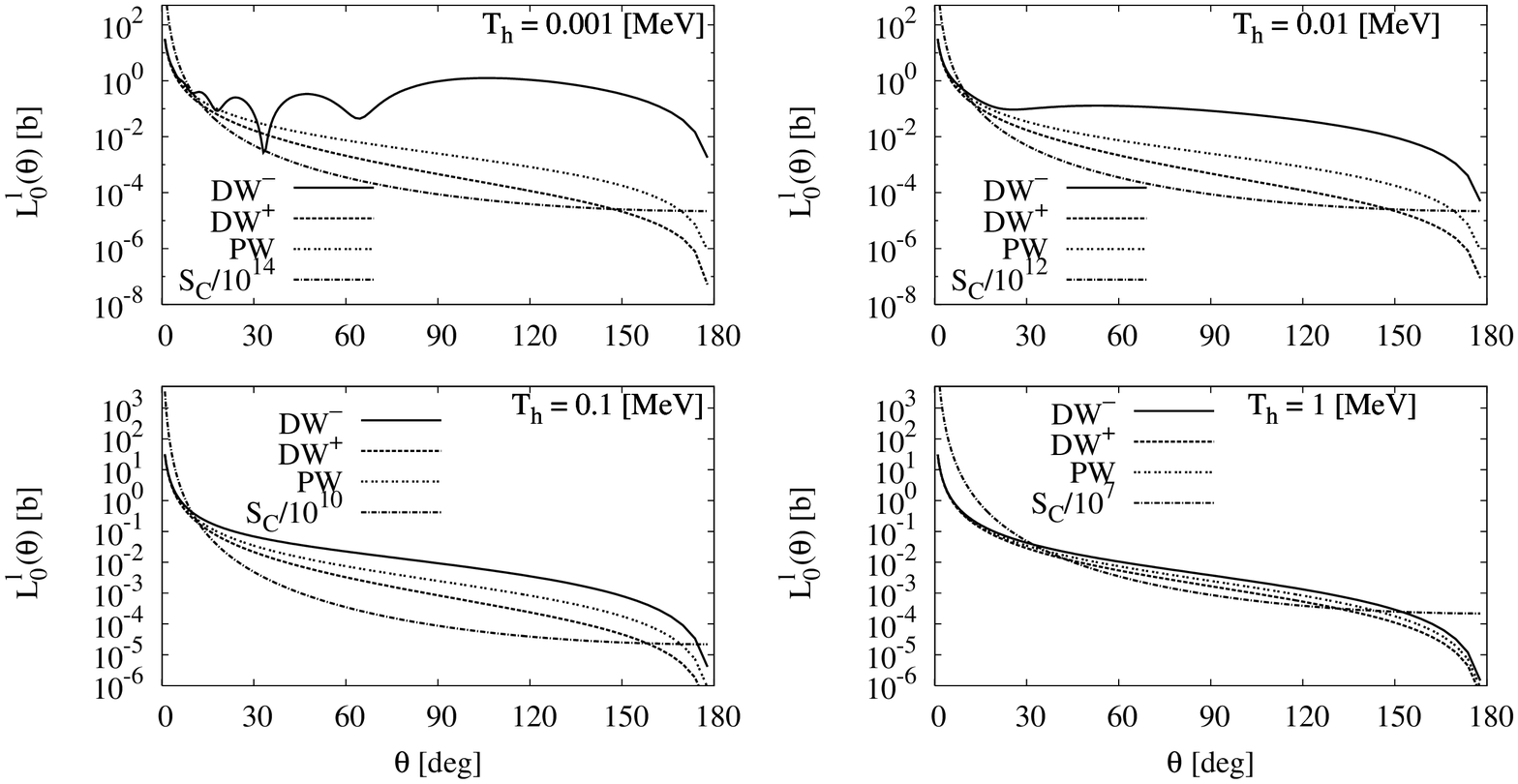}
\caption{The structure function $L^l_0(\theta)$ of the triple polarization
  differential cross section 
  in plane wave (PW) and distorted wave approximation
  for like charges (DW$^+$) and opposite charges (DW$^-$) 
 for various proton lab kinetic energies $T_h$. For comparison the Coulomb
 structure function $S_C(\theta)$ is shown in addition reduced by a
 factor 10$^{-n}$.}
\label{fig_L_0_l}

\includegraphics[width=1.\columnwidth]{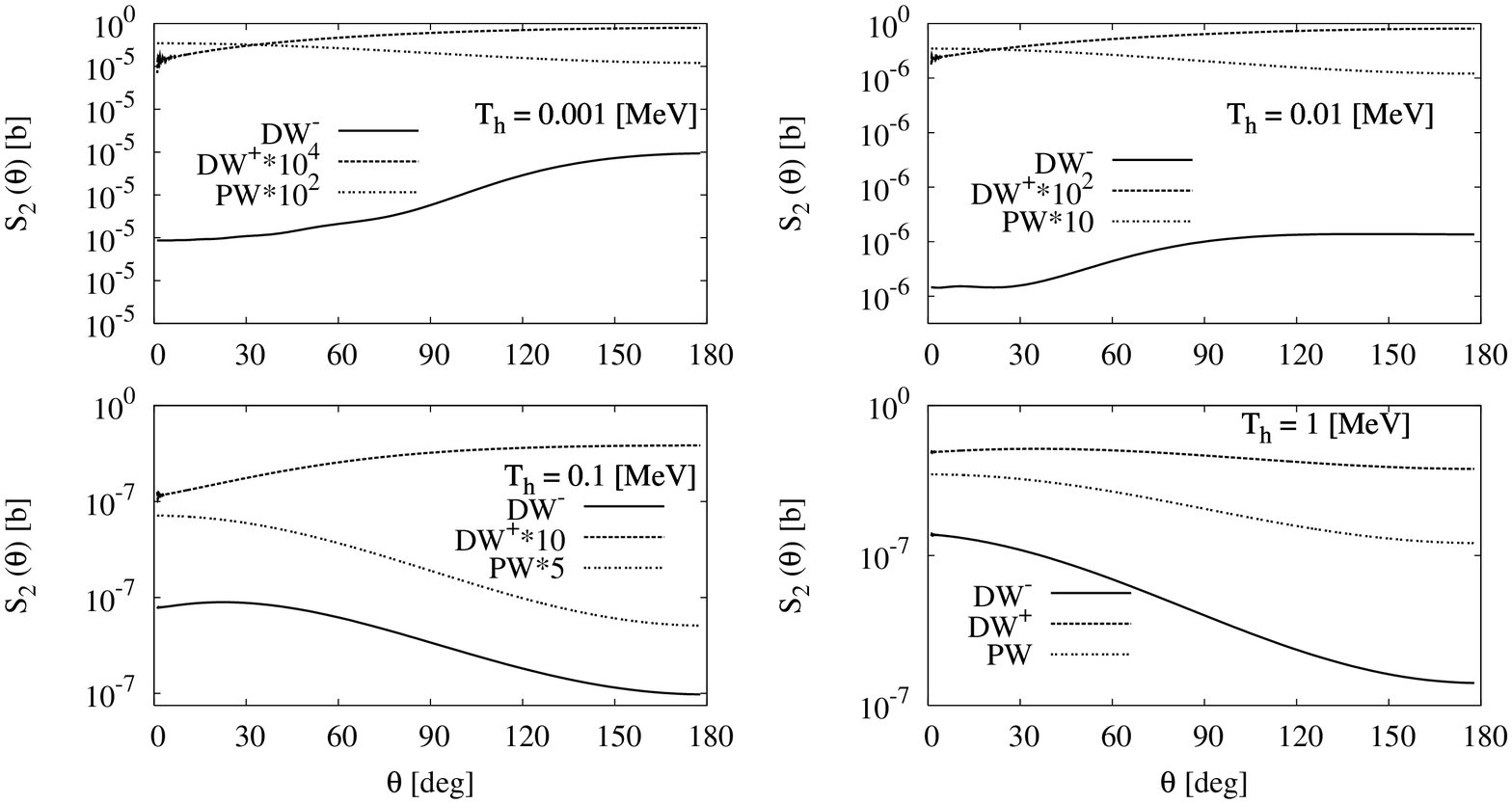}
\caption{The structure function $S_2(\theta)$ of the triple polarization
  differential cross section  
  in plane wave (PW) and distorted wave approximation
  for like charges (DW$^+$) and opposite charges (DW$^-$) 
 for various proton lab kinetic energies $T_h$. One should note the
 enhancement factors for PW and DW$^+$. }
\label{fig_S_2} 
\end{figure}

\begin{figure}[h]
\includegraphics[width=1.\columnwidth]{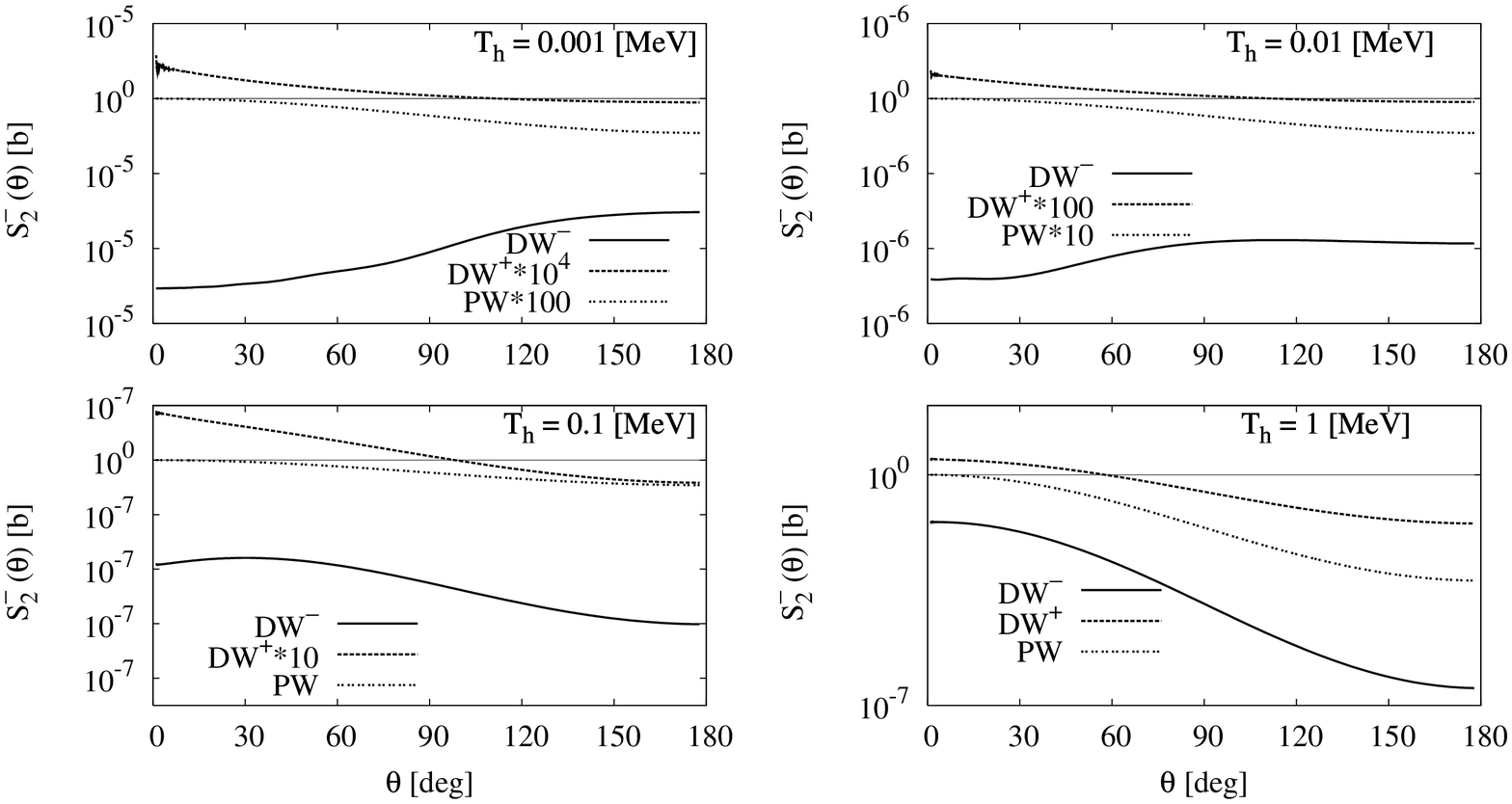}
\caption{The structure function $S_2^-(\theta)$ of the triple polarization
  differential cross section
  in plane wave (PW) and distorted wave approximation
  for like charges (DW$^+$) and opposite charges (DW$^-$) 
 for various proton lab kinetic energies $T_h$.  One should note the
 enhancement factors for PW and DW$^+$. }
\label{fig_S_2_minus}

\includegraphics[width=1.\columnwidth]{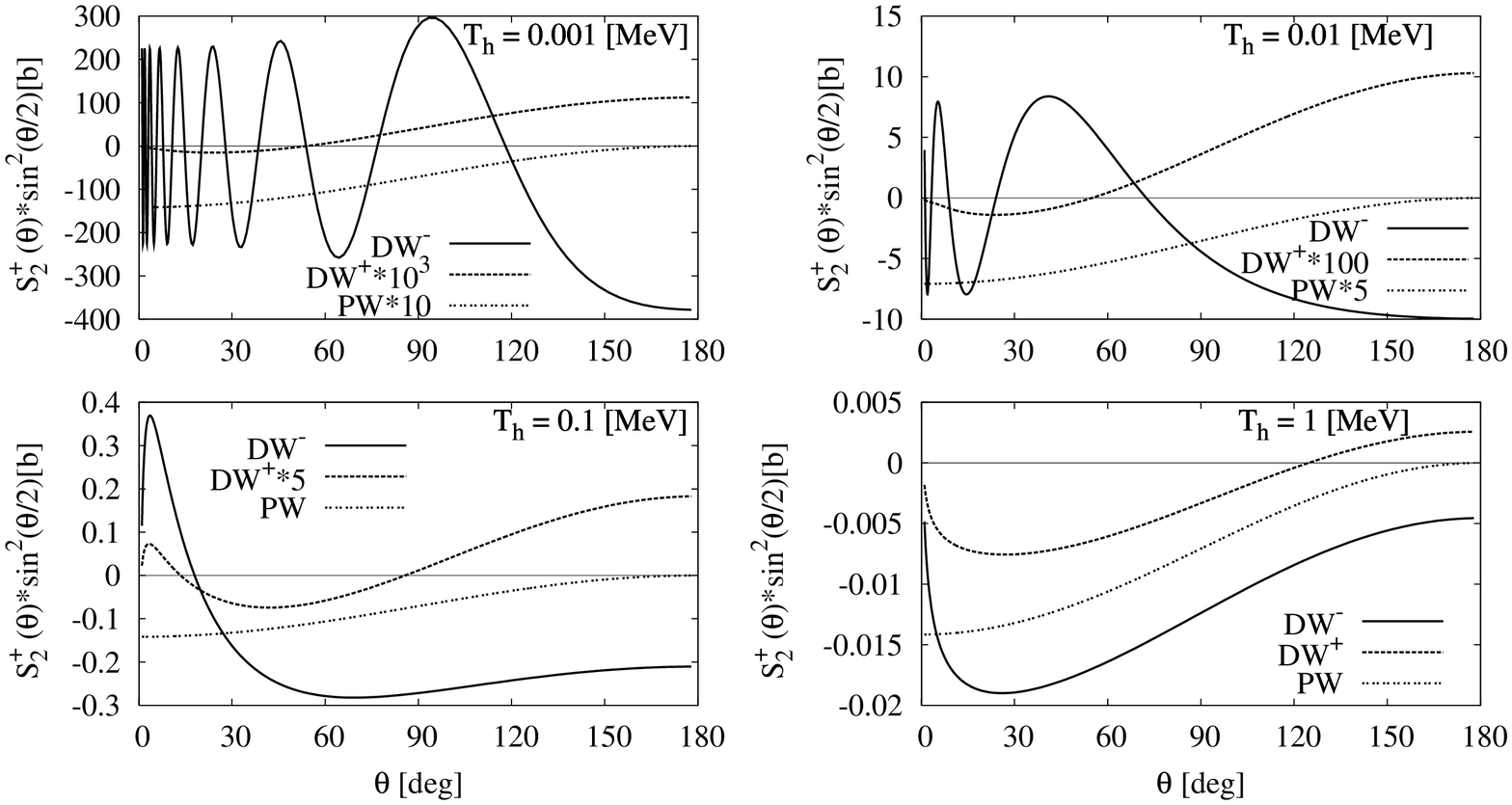}
\caption{The structure function $S_2^+(\theta)$ of the triple polarization
  differential cross section
  scattering multiplied by $\sin^2(\theta/2)$
  in plane wave (PW) and distorted wave approximation
  for like charges (DW$^+$) and opposite charges (DW$^-$) 
 for various proton lab kinetic energies $T_h$.  One should note the
 enhancement factors for PW and DW$^+$. }
\label{fig_S_2_plus} 
\end{figure}

\begin{figure}[h]
\includegraphics[width=1.\columnwidth]{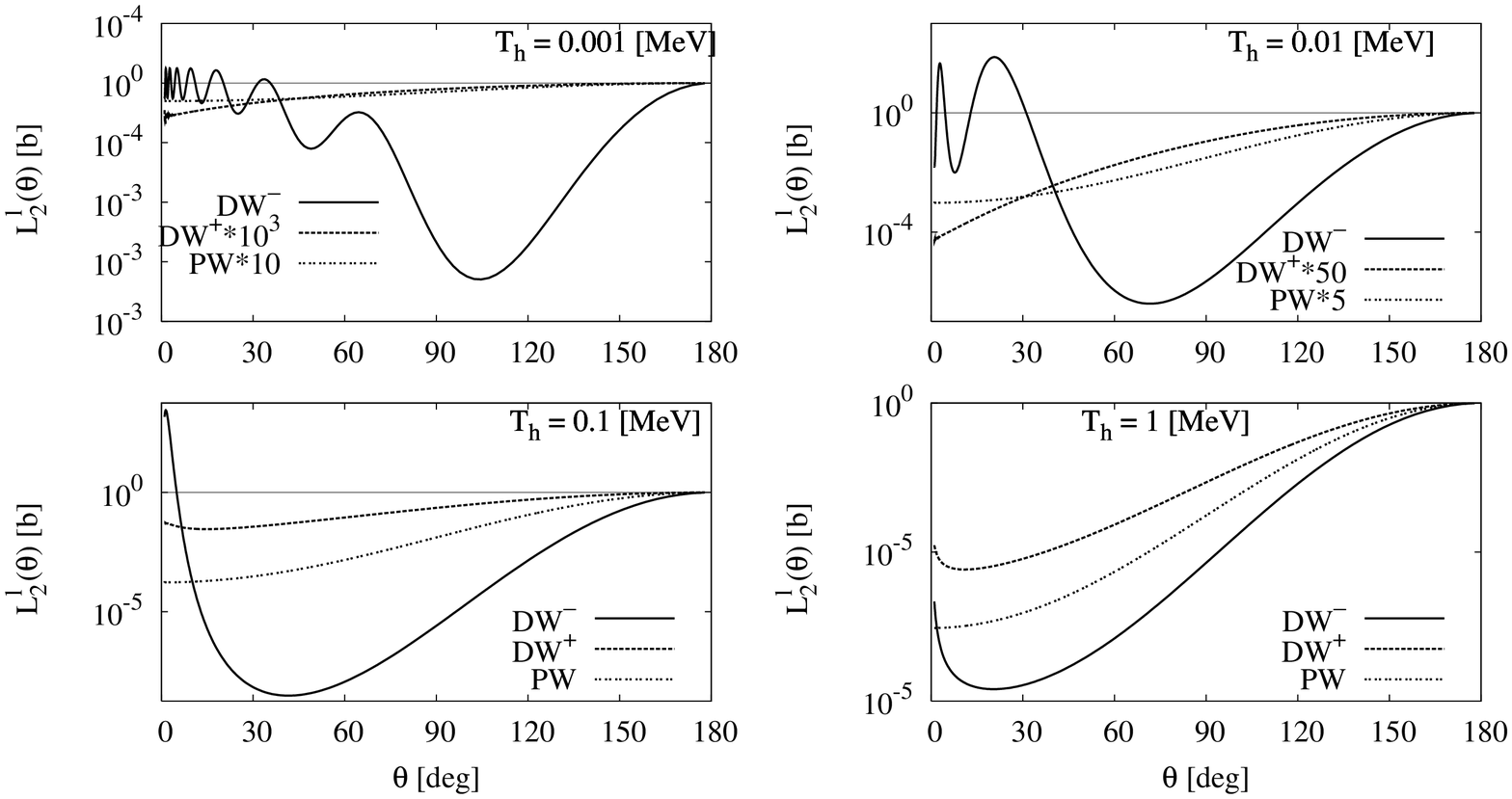}
\caption{The structure function $L_2^l(\theta)$ of the triple polarization
  differential cross section 
  in plane wave (PW) and distorted wave approximation
  for like charges (DW$^+$) and opposite charges (DW$^-$) 
 for various proton lab kinetic energies $T_h$. One should note the
 enhancement factors for PW and DW$^+$ in the upper panels. }
\label{fig_L_2_l} 
\end{figure}

 First, I will discuss the various structure functions
which determine the triple polarization cross section of
eq.~(\ref{hadron1}). For the plane wave approximation, one finds
easily the following expressions
\beqa
L_0^{l/h}(\theta)&=&-\frac{1}{4}\,V(c^{LS}_{l/h})^2\cot^2(\theta/2)\,,\\
L_2^{l/h}(\theta) &=& -V c^{LS}_{l/h}c^{SS}\cos^2(\theta/2)\,,\\
S_0(\theta)&=&2\,V (c^{SS})^2\,,\\
S_2^+(\theta)&=&-\frac{Vc^{SS}}{2p^2}\,\cot^2(\theta/2)\,,\\
S_2^-(\theta)&=&-2\,V (c^{SS})^2\sin^2(\theta/2)\,,\\
S_2(\theta)&=&-2\,V (c^{SS})^2(1+\sin^2(\theta/2))\,.
\eeqa
The structure functions, evaluated in the c.m.~frame
for several lab kinetic energies, are shown in Figs.~\ref{fig_S_0} through
\ref{fig_L_2_l} for the various approximations, i.e.\ plane wave
approximation (PW) and Coulomb distortion for like (DW$^+$) and
opposite charges (DW$^-$). 

The diagonal structure 
function $S_0$ in Fig.~\ref{fig_S_0}, induced by the hyperfine
interaction, shows a rather flat, almost constant angular
behaviour. Its size scales roughly proportional to the inverse of the
kinetic energy $T_h$. Compared to the plane wave approximation (PW),
the distorted wave approximation is strongly enhanced for unlike
charges (DW$^-$) and strongly suppressed for like charges (DW$^+$) by
several orders of magnitude 
for the lowest lab kinetic energy of $T_h=0.001$~MeV corrsponding to
$\eta_C\approx 5$. This enhancement resp.\ suppression is increasingly reduced
with growing kinetic energy $T_h$ and approaches the plane wave result
above $T_h=10$~MeV. The pure Coulomb
contribution $S_C$, however,  is much larger by more than ten orders of
magnitude. 

With respect to the other two diagonal structure 
functions from the leptonic and hadronic spin-orbit interactions
$L_0^{l}$ and $L_0^{h}$, it suffices to show only the former one in
Fig.~\ref{fig_L_0_l}, because $L_0^{h}$ differs in magnitude only
by the factor $(c_{LS}^h/c_{LS}^l)^2$ being smaller by about five
orders of magnitude. One readily notes that $L_0^{l}$
exhibits a strong peaking in the 
forward direction only and tends to oscillate at small angles for the
lowest $T_h$ considered here.  Over the whole angular range,
especially in the forward direction, $L_0^{l}$ is much 
larger by several orders of magnitude than $S_0$ but it is still
almost negligible compared to the size of $S_C$. The effect of
Coulomb distortion is qualitatively similar to what one observes in $S_0$.

Only these diagonal structure functions contribute to the
unpolarized cross section. However, as already mentioned, 
their relative contribution is
extremely small as can be seen by comparison with the pure Coulomb
structure function $S_C$ which is also shown in Fig.~\ref{fig_L_0_l}
indicated by the large reduction factor applied to $S_C$. 

The two hyperfine-hyperfine interference structure functions $S_2$ and
$S_2^-$, shown in Figs.~\ref{fig_S_2} and \ref{fig_S_2_minus}, which
both are negative througout, exhibit
a similar pattern, a smooth angular distribution with a slight
decrease in size at backward angles for the two lowest kinetic
energies, but with a slight increase for the two higher kinetic
energies like the PW result for all four cases. Again one notes
sizeable enhancements for opposite charges by Coulomb distortion 
and suppression for like charges for the lowest
energies $T_h$. The distortion effect decreases with increasing $T_h$.
Also these two structure functions are quite small like $S_0$ because
they are quadratic in the hyperfine amplitudes.

Much larger
is the third interference structure function $S_2^+$ because it is an
interference between the Coulomb and the hyperfine amplitudes. Thus it
is strongly forward peaked. For this reason, it is displayed in
Fig.~\ref{fig_S_2_plus} multiplied by $\sin^2(\theta/2)$. Moreover,
Coulomb distortion induces a strong oscillatory behavior, again in
conjunction with a large enhancement for opposite charges and strong
suppression for like charges. As expected, the distortion effect
diminishes with increasing kinetic energy $T_h$. I would like
to point out that in the corresponding Figs.~1 and 2 of
ref.~\cite{Are07} erroneously a factor $\sin^2(2\theta)$ instead of
the indicated factor $\sin^2(\theta/2)$ has been applied. 

Finally, the
spin-orbit-hyperfine interference structure function $L_2^l$ in
Fig.~\ref{fig_L_2_l} is comparable in size to $S_2$ and
$S_2^-$ but exhibits quite a different pattern. At the lowest
kinetic energy $T_h$ it is strongly enhanced by distortion for
opposite charges and
possesses a pronounced broad minimum around 100$^\circ$. It falls off
at forward and backward angles with many oscillations in the forward
direction. With increasing kinetic energy the minimum moves towards
smaller angles with fewer oscillations. Again the distortion effect
decreases strongly with increasing energy. Like the diagonal
structure function $L_0^h$, the hadronic interference structure function
$L_2^h$ is quite negligible. 

\begin{figure}[ht]
\includegraphics[width=1.\columnwidth]{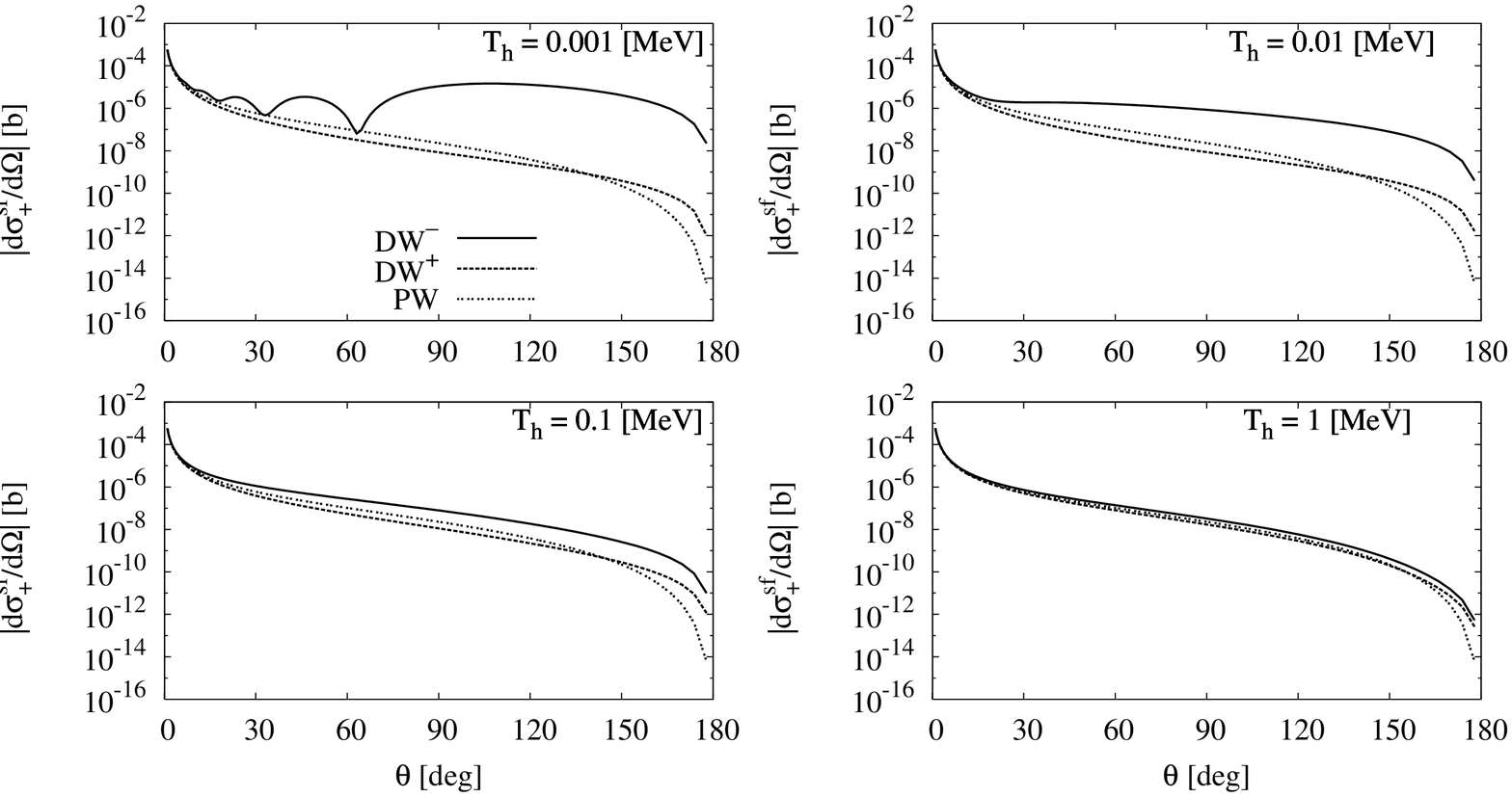}
\caption{Absolute value of spin-flip cross section $d\sigma_+^{sf}/d\Omega$ for
  initial hadron polarization parallel to lepton polarization along
  the initial relative momentum
  in plane wave (PW) and distorted wave approximation
  for like charges (DW$^+$) and opposite charges (DW$^-$) 
 for various proton lab kinetic energies $T_h$.}
\label{fig_spin_flip_plus}

\includegraphics[width=1.\columnwidth]{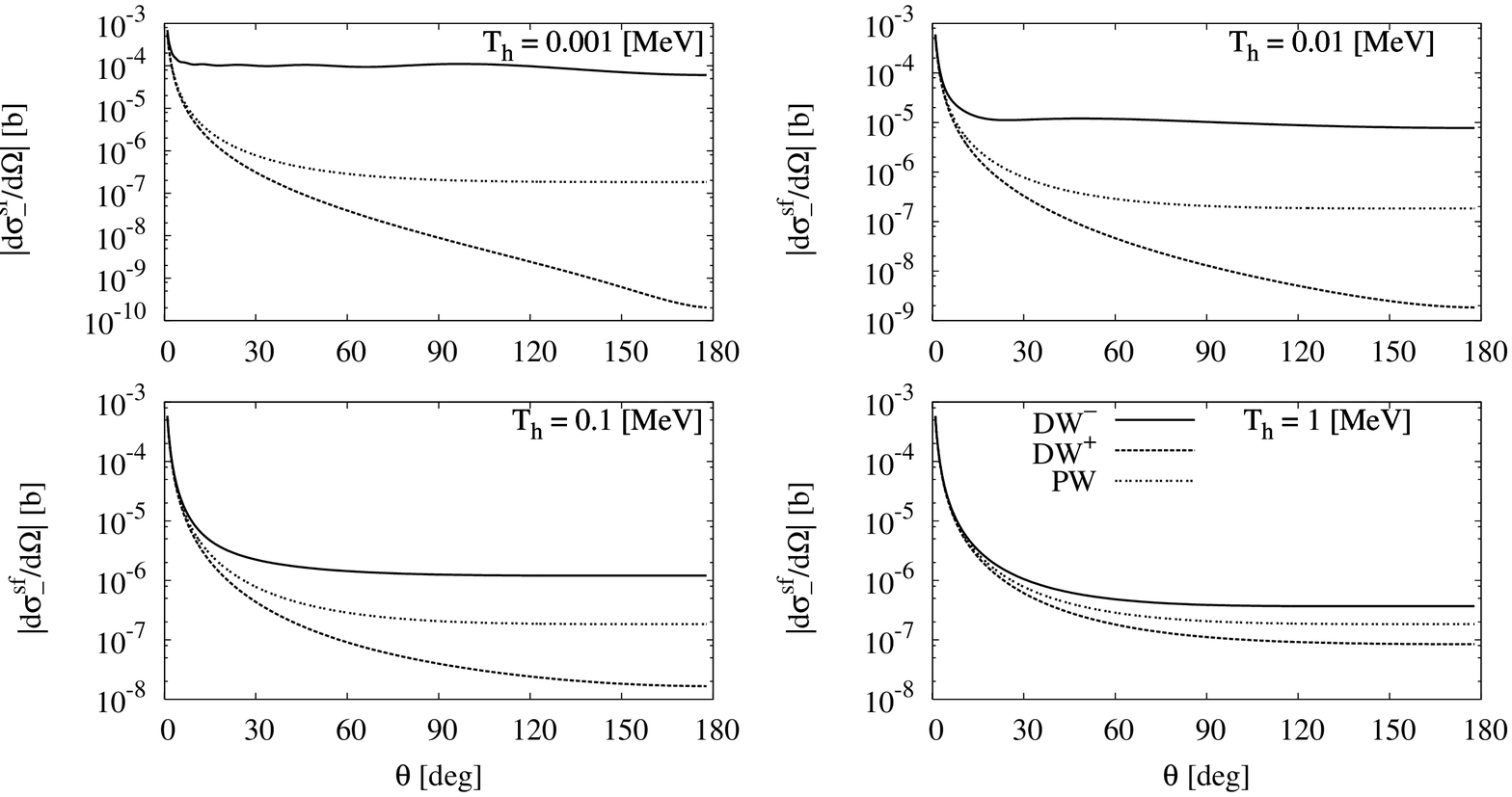}
\caption{Absolute value of spin-flip cross section $d\sigma_-^{sf}/d\Omega$ for
  initial hadron polarization opposite to lepton polarization along
  the initial relative momentum
  in plane wave (PW) and distorted wave approximation
  for like charges (DW$^+$) and opposite charges (DW$^-$) 
 for various proton lab kinetic energies $T_h$.}
\label{fig_spin_flip_minus} 
\end{figure}

\begin{figure}[ht]
\includegraphics[width=1.\columnwidth]{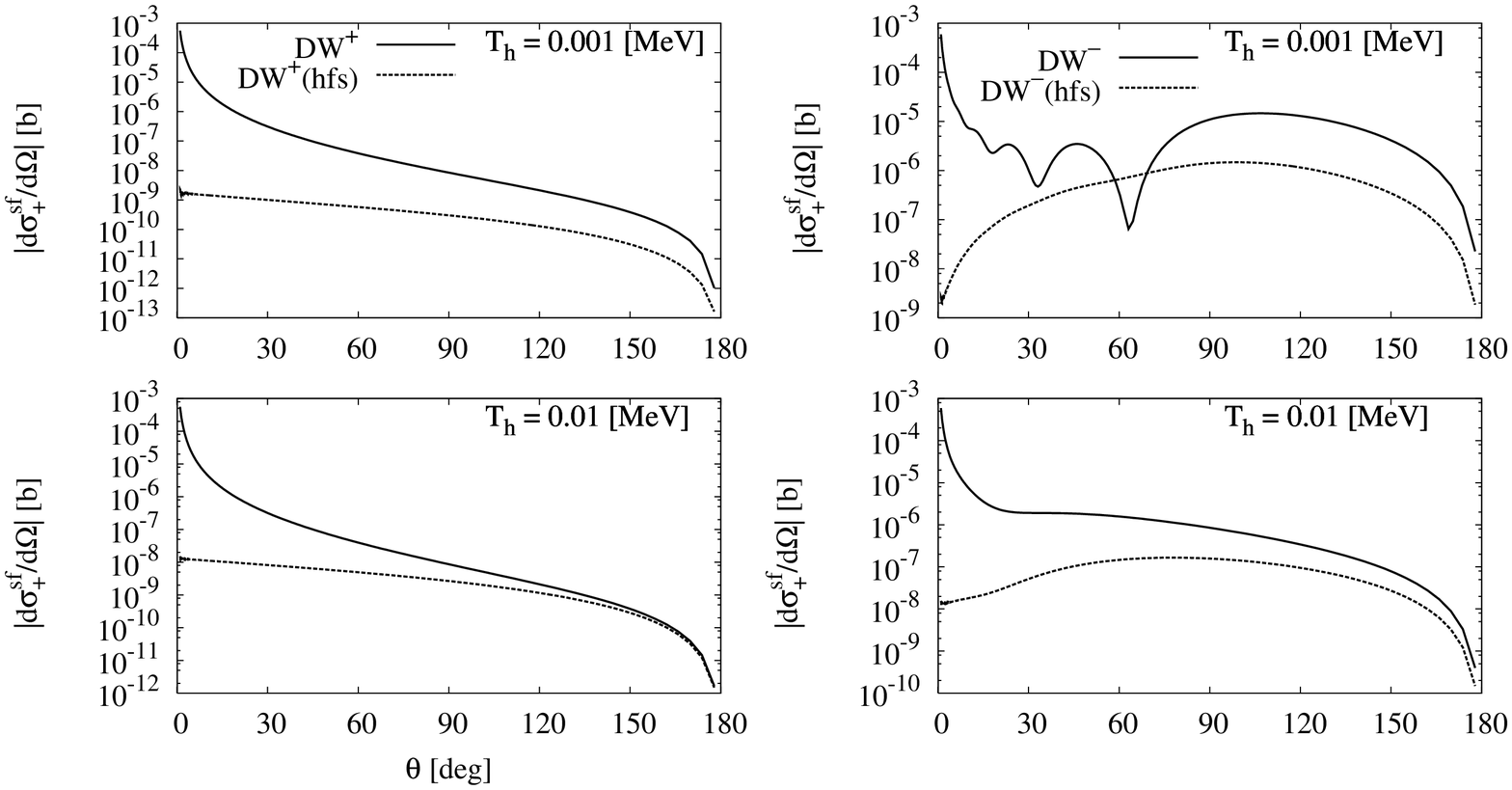}
\caption{Comparison of spin-flip cross section $d\sigma_+^{sf}/d\Omega$ for
  initial hadron polarization parallel to lepton polarization along
  the initial relative momentum with the hyperfine interaction alone
  (DW(hfs)) and with inclusion of the spin-orbit contribution (DW)
 for like charges (DW$^+$, left panels) and opposite charges (DW$^-$,
 right panels) for the two lowest proton lab kinetic energies $T_h$.}
\label{fig_spin_flip_plus_ss}

\includegraphics[width=1.\columnwidth]{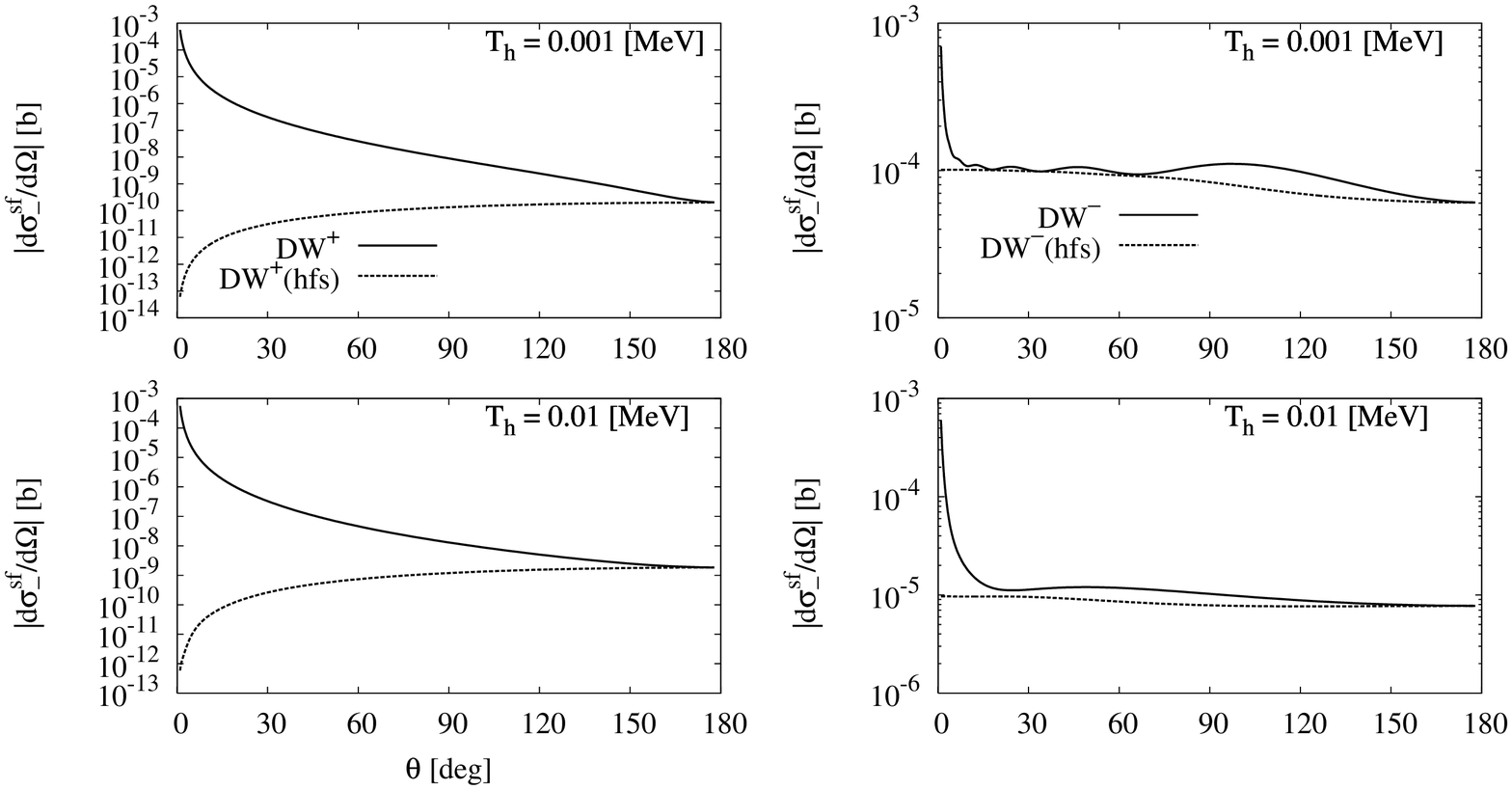}
\caption{Comparison of spin-flip cross section $d\sigma_+^{sf}/d\Omega$ for
  initial hadron polarization opposite to lepton polarization along
  the initial relative momentum with the hyperfine interaction alone
  (DW(hfs)) and with inclusion the of spin-orbit contribution (DW)
 for like charges (DW$^+$, left panels) and opposite charges (DW$^-$,
 right panels) for the two lowest proton lab kinetic energies $T_h$.}
\label{fig_spin_flip_minus_ss} 
\end{figure}

\subsection{The triple polarization cross section}

Now I will discuss the triple polarization cross section of
eq.~(\ref{hadron1}). Previously, in ref.~\cite{WaA07} only the
hyperfine amplitude besides the Coulomb one has been considered,
whereas the hadron spin-orbit interaction has 
been included already in ref.~\cite{MiS08}.  However, as 
mentioned above, its contribution to the helicity dependent part of
the spin-flip cross section in eq.~(\ref{hadron-flip}) is
negligible, whereas in the helicity independent part the diagonal contribution $L_0^h$ is
comparable in size to $S_2$ in the forward direction (compare
Figs.~\ref{fig_L_0_l} and \ref{fig_S_2}). Much more important
is the leptonic spin-orbit contribution which, however, appears in the
non-spin-flip cross section only (see eq.~(\ref{hadron-nonflip})) where
it is buried completely by the Coulomb contribution $S_C$.

The results for the spin-flip cross section for parallel initial spin orientations of
hadron and lepton is shown in Fig.~\ref{fig_spin_flip_plus} while the
one for the opposite spin orientation in
Fig.~\ref{fig_spin_flip_minus}. One notes again the strong influence
of Coulomb distortion. Furthermore, the leptonic spin-orbit
interaction plays a relatively important role in the region of the
minimum as can be seen in
Figs.~\ref{fig_spin_flip_plus_ss} and \ref{fig_spin_flip_minus_ss}
where a comparison is exhibited with the case for which the spin-orbit
interaction is switched off (curves labeled ``(hfs)''). One readily notes a substanial increase
when the spin-orbit part is included compared to the pure hyperfine
case.  Furthermore, at the lowest two energies
the spin-orbit interaction induces oscillations, in particular in the
forward direction. The difference of the two spin-flip cross sections
determines the net hadron polarization in a storage ring of initially
unpolarized hadrons scattered at polarized leptons. 

\begin{figure}[ht]
\includegraphics[width=1.\columnwidth]{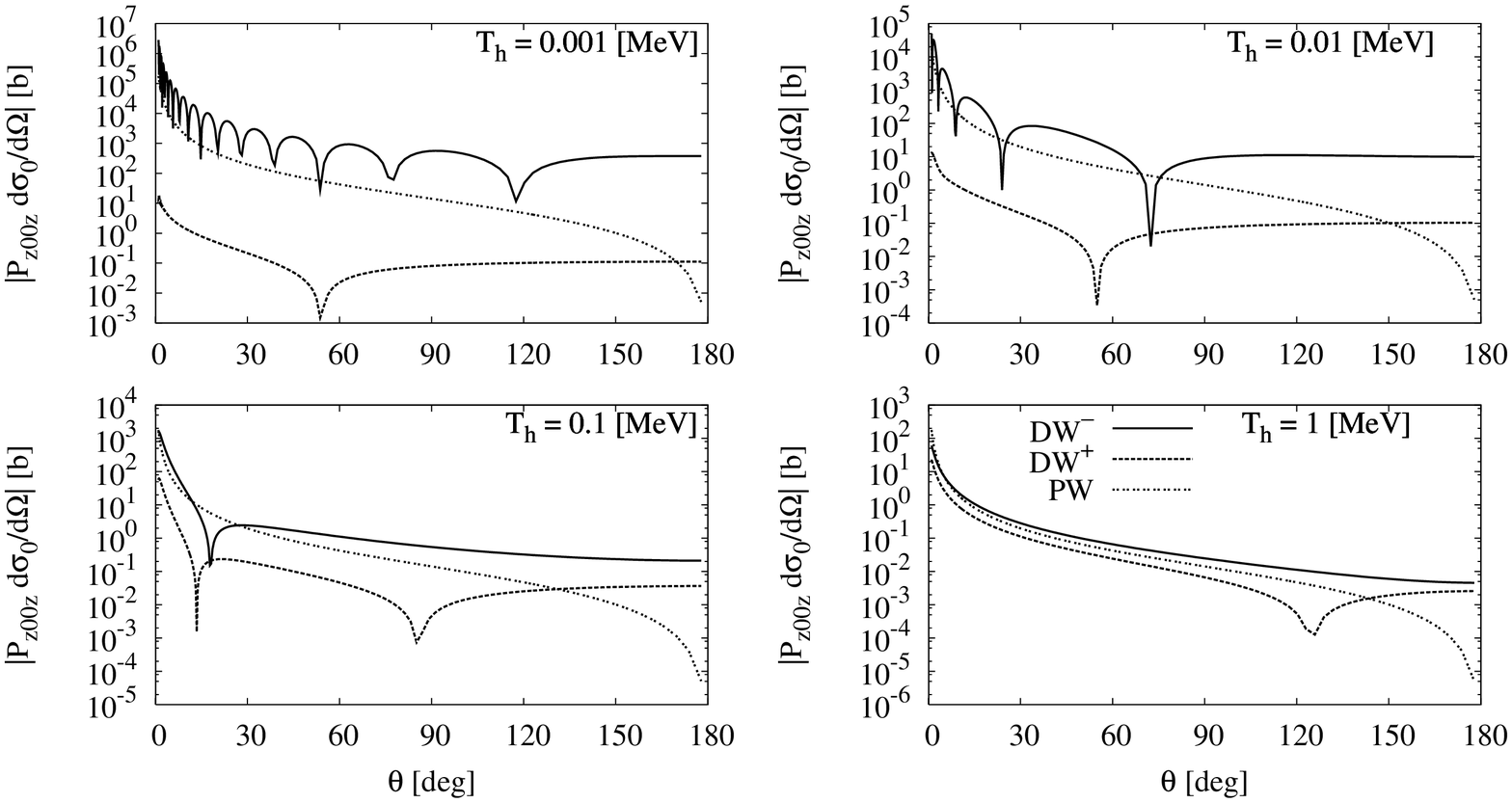}
\caption{Lepton-hadron polarization transfer cross section $P_{z00z} d\sigma_0/d\Omega$
 with inclusion of spin-orbit contribution 
 for like charges (DW$^+$) and opposite charges (DW$^-$) for several
 proton lab kinetic energies $T_h$. } 
\label{fig_pzz_plus_minus}

\includegraphics[width=1.\columnwidth]{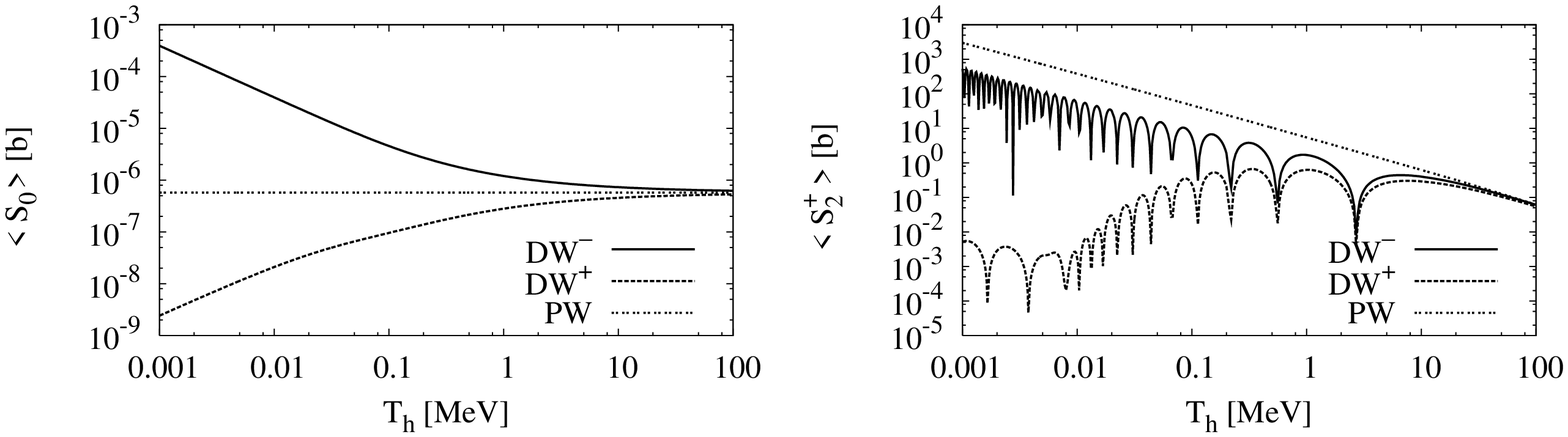}
\caption{The integrated structure functions $\langle S_0 \rangle$  (left panel)
  and $\langle  S_2^+ \rangle$  (right panel) as function of the proton lab kinetic
  energy $T_h$ for plane wave approximation (PW) and with Coulomb
  distortion for like charges (DW$^+$) and opposite charges (DW$^-$).} 
\label{Fig_int_s2plus}

\includegraphics[width=1.\columnwidth]{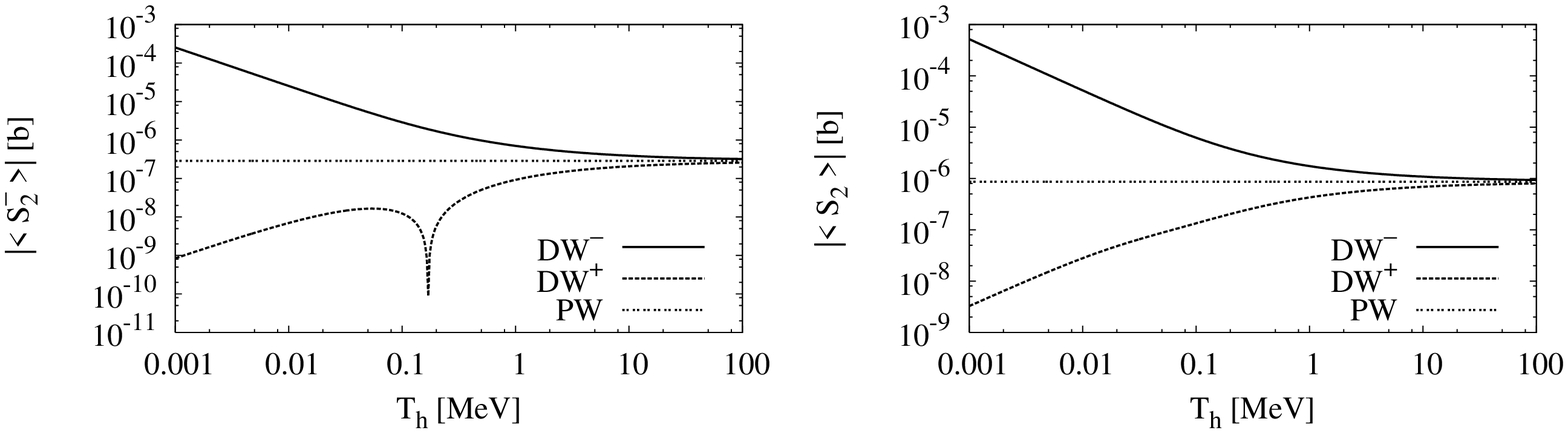}
\caption{The integrated structure functions $\langle S_2^- \rangle$  (left panel)
  and $\langle S_2 \rangle$  (right panel) as function of the proton lab kinetic
  energy $T_h$ for plane wave approximation (PW) and with Coulomb
  distortion for like charges (DW$^+$) and opposite charges (DW$^-$).} 
\label{Fig_int_sminus} 
\end{figure}

The non-spin-flip cross section in eq.~(\ref{hadron-nonflip}) is
overwhelmingly dominated by the Coulomb contribution $S_C$. 
The small dependence on $\lambda_l^i$ and $\lambda_h$ leads to
different scattering strengths for the hadron polarization parallel or
antiparallel to the lepton polarization (see
eq.~(\ref{diff-nonflip})). The resulting lepton-hadron polarization
transfer $P_{z00z}$ is shown in Fig.~\ref{fig_pzz_plus_minus}. The
dominance of the hyperfine amplitude is clearly seen. 

\begin{figure}[ht]
\includegraphics[width=1.\columnwidth]{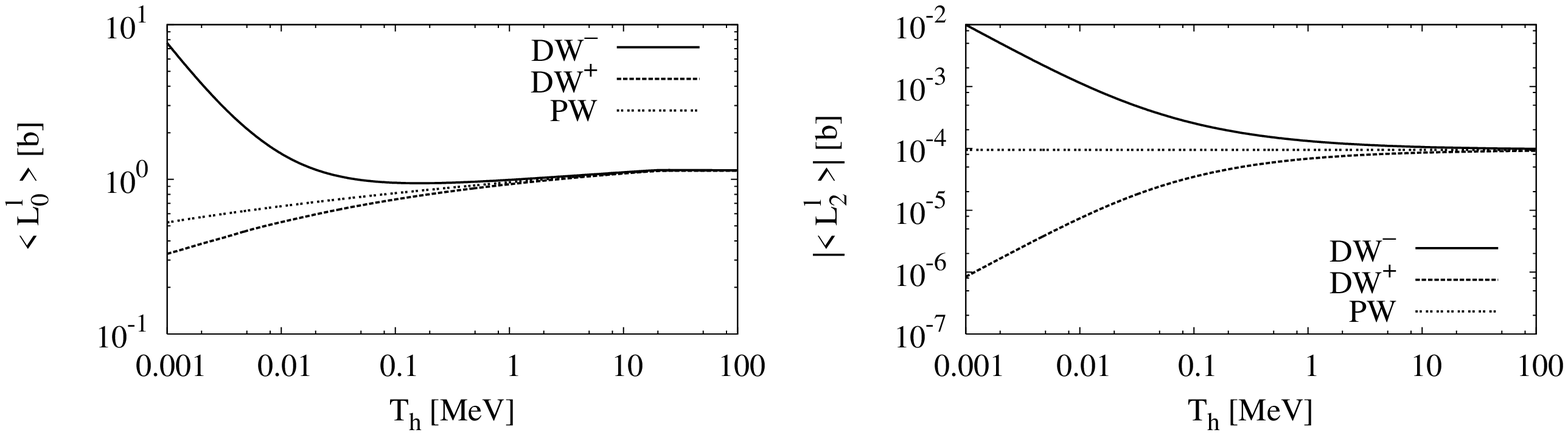}
\caption{The integrated structure functions $\langle L_0^l \rangle$  (left panel)
  and $\langle  L_2^l \rangle$  (right panel) as function of the proton lab kinetic
  energy $T_h$ for plane wave approximation (PW) and with Coulomb
  distortion for like charges (DW$^+$) and opposite charges (DW$^-$).} 
\label{Fig_int_L}

\includegraphics[width=1.\columnwidth]{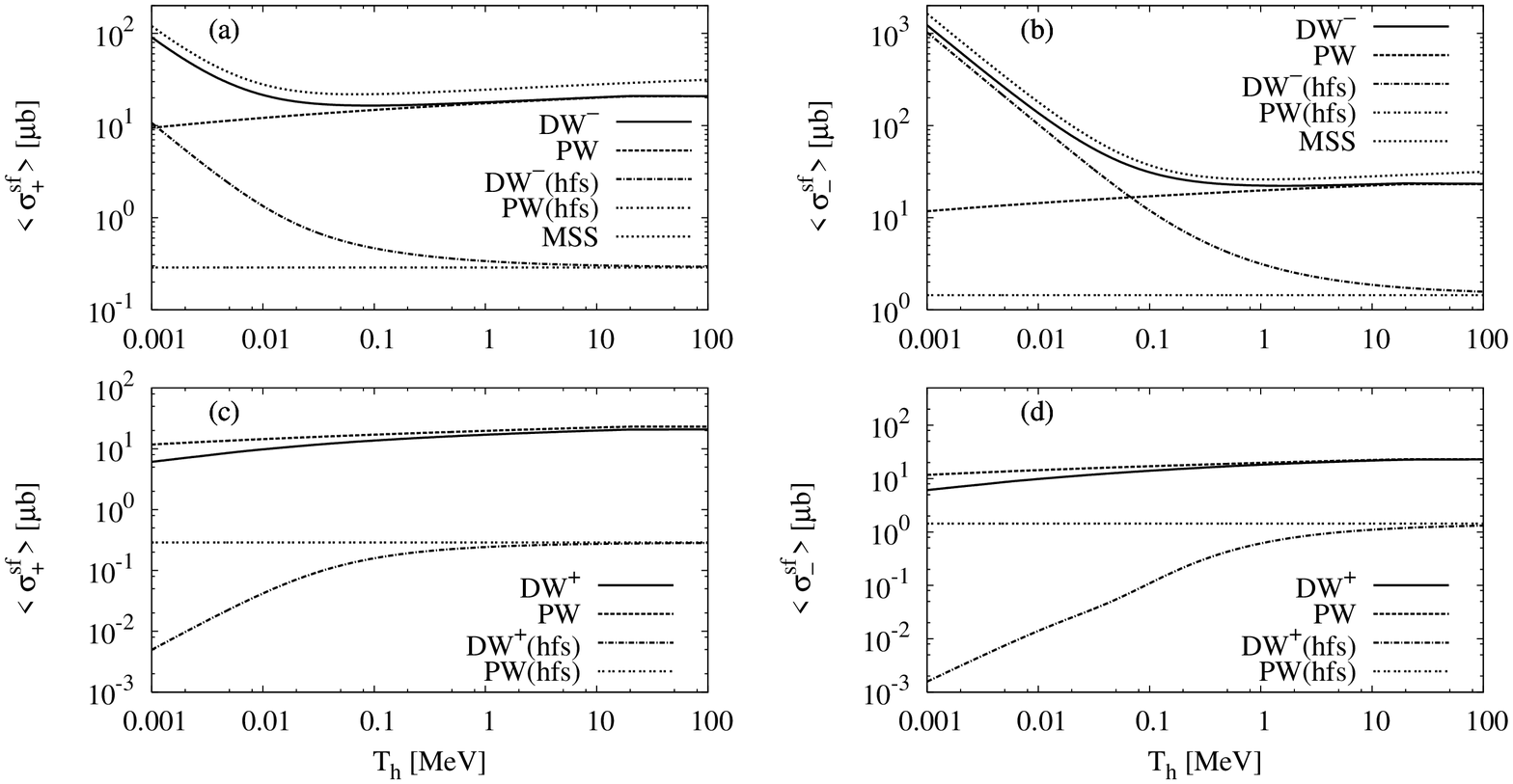}
\caption{The integrated spin-flip cross section $\langle
  {\sigma^{sf}_{+}} \rangle$  (left panels) 
  and $\langle {\sigma^{sf}_{-}} \rangle$
  (right panels) as function of the proton lab kinetic 
  energy $T_h$ for plane wave approximation (PW) and with Coulomb
  distortion for like charges (DW$^+$, lower panels) and opposite charges
  (DW$^-$, upper panels). Curves labeled ``hfs'' include the hyperfine
  amplitude only, and in the upper panels the curves labeled
  ``MSS'' represent the results of ref.~\cite{MiS08}.} 
\label{Fig_int_spin-flip_plus} 
\end{figure}

\begin{figure}[ht]
\includegraphics[width=1.\columnwidth]{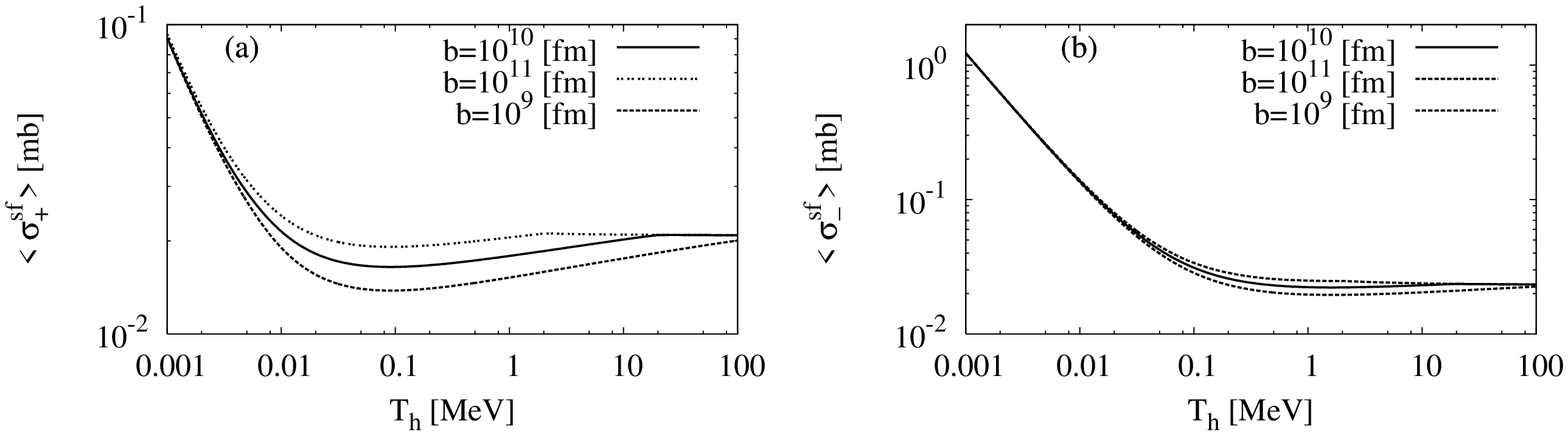}
\caption{Dependence of the integrated spin-flip cross sections for
  opposite charges on the regularization parameter $b$.
}
\label{Fig_int_spin-flip-b-dep} 
\end{figure}

\subsection{The integrated structure functions and cross sections}

Finally, I will present results for the integrated
structure functions and spin-flip cross sections which are the
relevant quantities for the polarization build-up in a storage
ring. They are defined by the integration over the solid angle except
for the small cone in the forward direction with $\theta < \theta_{min}$,
where the minimal scattering angle is defined by the requirement that
the impact parameter should not exceed a given value b
\beq
\theta_{min}=2\arctan(\eta_C/l)\,,
\eeq
with $l=bp$ as classical angular momentum. In the present work I have
chosen $b=10^{10}$. The choice of this value has been justified in
ref.~\cite{WaA07}. The dependence on this parameter is discussed
below. Thus for any structure
function or cross section ${\cal O}(\theta)$ I define
\beq
\langle {\cal O} \rangle= 2\pi\int_{\theta_{min}}^\pi d(\cos \theta)
{\cal O}(\theta)\,.
\eeq

For the plane wave approximation, one finds
easily the following expressions
\beqa
\langle  L_0^{l/h}(\theta) \rangle&=&-2\pi\,V(c^{LS}_{l/h})^2
\Big(\ln(\sin(\theta_{min}/2))+\frac{1}{2}\cos^2(\theta_{min}/2)\Big) \,,\\
\langle  L_2^{l/h}(\theta) \rangle&=& -2\pi\,V c^{LS}_{l/h}c^{SS}
\Big(1-\sin^2(\theta_{min}/2)(1+\cos^2(\theta_{min}/2))\Big)
\,,\\
\langle  S_0(\theta) \rangle&=&8\pi\,V (c^{SS})^2\,,\\
\langle  S_2^+(\theta) \rangle&=&4\pi\,V \frac{c^{SS}}{p^2}
\Big(\ln(\sin(\theta_{min}/2))-\frac{1}{2}\cos^2(\theta_{min}/2)\Big) 
\,,\\
\langle  S_2^-(\theta) \rangle&=&-4\pi\,V (c^{SS})^2
\cos^2(\theta_{min}/2)(1+\sin^2(\theta_{min}/2)) 
\,,\\
\langle  S_2(\theta) \rangle&=&-4\pi\,V (c^{SS})^2(3-\sin^4(\theta_{min}/2))\,.
\eeqa
Thus in plane wave all of the integrated structure functions except
for $\langle S_2^+ \rangle$ are almost independent of $T_h$ except for a very weak
dependence via the minimal scattering angle. One should note the
logarithmic divergence for $\theta_{min} \to 0$ in $\langle
L_0^{l/h} \rangle$ and $\langle  S_2^+ \rangle$. It
corresponds to the logarithmic divergence in the angular momentum $l$
of the partial wave expansion noted in
ref.~\cite{MiS08} which appears when integrating over the whole
range of scattering angles. 

The results are exhibited in Figs.~\ref{Fig_int_s2plus} through
\ref{Fig_int_L} for the structure functions and in
Fig.~\ref{Fig_int_spin-flip_plus} for the spin-flip cross
sections. The integrated structure functions show a strong increasing
influence of Coulomb distortion with decreasing hadron kinetic energy
$T_h$ leading to large enhancements for opposite charges and strong
suppression for like charges compared to the plane wave case. The only
exception is $\langle S_2^+ \rangle$ for which Coulomb distortion results in a
reduction for both cases, however much stronger for like charges. The
reason for this feature is the strong oscillatory behavior in the
angular distribution of $\langle S_2^+ \rangle$ at low $T_h$ (see
Fig.~\ref{fig_S_2_plus} ). 

The integrated spin-flip cross section is given by
\beq
\langle
  {\sigma^{sf}_{\pm}} \rangle=
2 \langle L_0^{h}  \rangle -\langle S_2 \rangle 
+2\pm\Big[\langle S_2^-\rangle+\langle L_2^{h} \rangle \Big]
\,.\label{int-hadron-flip}
\eeq
The corresponding integrated spin flip cross section of Milstein et
al.~\cite{MiS08} reads according to their eq.~(21)  
\beq
\langle
  {\sigma^{sf}_{\pm}} \rangle_{MSS}=\pi\Big(\frac{\alpha\mu_p}{M_p}\Big)^2
\Big[(2\pi\eta_C)^2(\frac{11}{6}-\ln 2)+\ln(l_{max}/\eta_C)^2\mp (2\pi\eta_C)^2 \Big]\,.
\eeq
The appearing logarithmic divergence in the angular momentum $l$
is regularized by choosing a finite
$l_{max}$ determined by the classical relation $l_{max}=bp$ which
corresponds to the choice of a minimum scattering angle in the present
work. The integrated strength of the spin-flip cross sections in
Fig.~\ref{Fig_int_spin-flip_plus} 
shows as expected with decreasing $T_h$ a growing strong influence
of Coulomb effects via the hyperfine and  hadronic spin-orbit
interactions. The latter is only important in the spin-independent part
of the spin-flip cross section while its influence in the
spin-dependent part is negligible. The results of Milstein et
al.~\cite{MiS08}, shown in the upper
panels of Fig.~\ref{Fig_int_spin-flip_plus} for opposite charges, are
comparable to our results but display a slight overestimation which is
probably due to different approximations in ~\cite{MiS08}. The
dependence of the integrated spin-flip cross 
section for opposite charges on the regularization parameter $b$
is exhibited in Fig.~\ref{Fig_int_spin-flip-b-dep} for $b=10^9$, $10^{10}$ and
$10^{11}$~fm. It appears to be quite weak. 

\begin{figure}[h]
\includegraphics[width=.7\columnwidth]{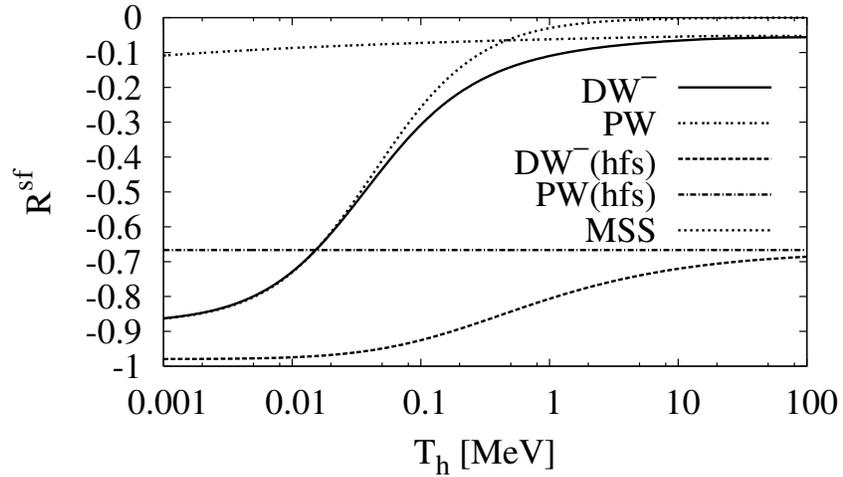}
\caption{Ratio of the spin dependent part of the spin-flip cross
  section over its spin-independent part as function of the proton lab kinetic 
  energy $T_h$ for plane wave approximation (PW) and with Coulomb
  distortion for opposite charges  (DW$^-$): present calculation and
  the result of ref.~\cite{MiS08} (MSS). For the curves labeled
  (hfs)  only the hyperfine amplitude is  included. }
\label{Fig_int_spin-flip_ratio} 
\end{figure}

The relevant quantity for a
polarization build-up in a storage ring is the ratio of the
spin-independent part over the spin-dependent part  
\beq
R^{sf}=2\frac{\langle S_2^-\rangle+\langle L_2^{h}\rangle}
{2 \langle L_0^{h} \rangle -\langle S_2 \rangle }\,,
\eeq
which is shown in Fig.~\ref{Fig_int_spin-flip_ratio} . One readily
notes the reduction of this ratio by the hadronic spin-orbit
interaction, in particular quite strong at higher energies but only
about 12~\% at the lowest energy. This fact clearly shows the
importance of the hadronic spin-orbit interaction besides the
hyperfine contribution. 

\begin{figure}[h]
\includegraphics[width=1.\columnwidth]{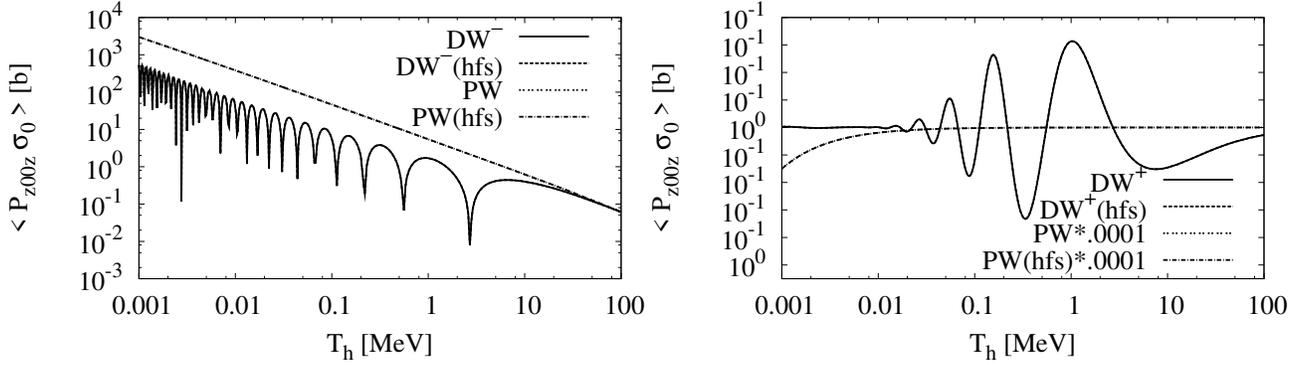}
\caption{The integrated polarization transfer cross section $\langle
  P_{z00z} \sigma_0 \rangle$  
 as function of the proton lab kinetic 
  energy $T_h$ for plane wave approximation (PW) and with Coulomb
  distortion for like charges (DW$^+$, right panel) and opposite charges
  (DW$^-$, left panel). For the curves labeled (hfs)  only the hyperfine amplitude is
  included. } 
\label{Fig_int_Pzz} 
\end{figure}

Finally, I show for completeness in Fig.~\ref{Fig_int_Pzz} the
integrated polarization transfer cross section $\langle
  P_{z00z} \sigma_0 \rangle$. It is dominated by the hyperfine
  interaction whereas the spin-orbit contribution is negligible. 

\section{Conclusions }
Formal expressions for polarization observables in electromagnetic
hadron-lepton scattering have been derived within a nonrelativistic
framework including the central Coulomb force as well as the lepton
and hadron spin-orbit and  hyperfine interactions. The latter have
been treated in a distorted wave approximation. 
Special emphasis has been 
laid on the triple polarization cross section with polarizations of
the initial hadron and lepton and of the final
hadron along the incoming hadron momentum. The structure functions
which determine the differential 
triple polarization cross section have been evaluated in plane and
distorted wave approximations for hadron lab kinetic energies between
1 keV and 100 MeV. 

For the evaluation of the Coulomb distortion two
different methods have been employed: (i) an integral representation
of the nonrelativistic Coulomb scattering wave function and (ii) a
partial wave expansion. These two independent methods have served as a
mutual check for the numerical accuracy of the results. 

As expected, the distortion effects are very important  at low
energies in the small polarization observables, which are driven by
the spin-orbit and hyperfine interactions,
leading to sizeable enhancements for opposite charges and suppressions
for like charges according to the Coulomb attraction resp.\
repulsion. This is shown in detail for the structure functions of the
triple polarization cross section and for the special case of the
spin-flip differential cross section.

The leptonic spin-orbit interactions plays a
minor role in the non-spin-flip cross section in its spin-dependent
part, which, however, as a whole is by many orders of magnitude
smaller than the spin-independent part, dominated by the Coulomb term
$S_C$.  The influence of the spin-orbit and hyperfine interactions
on the unpolarized cross section is almost negligible in the whole
range of energies studied here. 

With respect to the integrated spin-flip cross sections our previous
work has been extended by the inclusion of the hadronic spin-orbit
interaction which shows a non-negligible effect in the
spin-independent part changing sizeably the ratio of the integrated
strength of its spin-dependent over the spin-independent part.

\section*{Acknowledgment}
I would like to thank Thomas Walcher for his continued interest in
this work and for many useful discussions. This work has been
supported by the SFB 443 of the Deutsche Forschungsgemeinschaft.  

\appendix
\renewcommand{\theequation}{A.\arabic{equation}}
\setcounter{equation}{0}
\renewcommand{\thefigure}{A\arabic{figure}}
\setcounter{figure}{0}

\section*{Appendix: Evaluation of hyperfine and 
 spin-orbit  interaction}
For the evaluation of the Coulomb distortion of the amplitude $\mathbf b$
of the spin-orbit interaction in DWBA as given in eq.~(\ref{ls_dwba}) and
the tensor amplitudes 
$d^{[2]}$ of the hyperfine interaction in eq.~(\ref{t_dwba_2}) two
different method have been applied: (i) An integral
representation and (ii)
a partial wave expansion of the Coulomb wave function. 
For convenience I set in this appendix $\eta=\eta_C$.

\subsection{Integral representation}
In~\cite{Are07} a detailed description of this method for the
evaluation of $d^{[2]}$  has been
given. Therefore, I will only summarize the result. 
The method is based
on an integral representation of the 
confluent hypergeometric function as proposed in~\cite{LeA01}
\beqa
_1F_1(-i\eta,1;ix)&=& Q(\eta)\int_0^1 dt\,f(t,\eta)\,
(1+t\frac{\partial}{\partial t})\,e^{ix(1-t)}\,,
\label{intrep}
\eeqa
with
\beq
Q(\eta)=\frac{\sinh{\pi\eta}}{\pi\eta},\,\, \mbox{and}\,\, f(t,\eta)=
e^{i\eta\ln{\frac{t}{1-t}}}\,.
\eeq
With the help of this representation, the hyperfine tensor
$d^{[2]}_{ij}$ and the spin orbit vector $\mathbf b$ are expressed as
two-dimensional integrals, as described below.

\subsubsection*{Hyperfine interaction}
For the hyperfine tensor one finds
\beqa
d^{[2]}_{ij}(\eta,\theta)&=&c^{SS}\,N(\eta)
\Big[\widetilde{d}^{\,[2]}_{ij}(1 ,\eta,\theta) 
+i\eta\int_0^1\frac{dt}{1-t}\,e^{-i\eta\ln{(1-t)}}
\Big(\widetilde{d}^{\,[2]}_{ij}(1 ,\eta,\theta)-e^{i\eta\ln{t}}
\widetilde{d}^{\,[2]}_{ij}(t ,\eta,\theta)\Big)\Big]\,,\label{dij}
\eeqa
where 
\beq
N(\eta)=e^{-\pi\eta}\,\frac{\sinh{(\pi\eta)}}{\pi\eta}
\eeq
is a normalization factor and
\beqa
\widetilde{d}^{\,[2]}_{ij}(t, \eta,\theta)&=&A_{ij}(t,1
,\eta,\theta)\,I_{SS}(c(t,1))\nonumber \\&&
+i\eta\int_0^1\frac{dt'}{1-t'}\,e^{-i\eta\ln{(1-t')}}
\Big(A_{ij}(t,1)\,I(c(t,1))
-e^{i\eta\ln{t'}}A_{ij}(t,t')\,I_{SS}(c(t,t'))\Big)\,.\label{dtildeij}
\eeqa
Here I have introduced the tensor
\beqa
A_{ij}(t,t')&=&3\hat{a}_i(t,t')\,\hat{a}_j(t,t')-\delta_{ij} \,,
\eeqa
with
\beq
\hat a(t,t')=\frac{\mathbf p\,t-\mathbf p^{\,\prime}\,t'}{pg(t,t')} \,,
\eeq
and
\beqa
g(t,t')&=&[t^2+t'^2-2tt'\cos{\theta}]^{1/2}\,.\label{gtts}
\eeqa
Furthermore, $I_{SS}(c)$ denotes the integral
\beqa
I_{SS}(c)&=& \int_0^\infty\frac{dx}{x}\,e^{icx}\,j_2(x)\nonumber\\
&=&\frac{1}{3}
-\frac{1}{2}\,c^2-\frac{1}{4}\,c(1-c^2)\,
\Big(\ln\Big|\frac{c+1}{c-1}\Big|-i\pi\,\Theta(1-c)\Big)\,,\label{IntSS}
\eeqa
with $\Theta(x)$ as the Heaviside step function and 
\beqa
c(t,t')&=&\frac{2-t-t'}{g(t,t')}\,.
\eeqa
One should note that ${d}^{[2]}_{ij}$ and
$\widetilde{d}^{\,[2]}_{ij}(t)$ are functions in  $\theta$ and $\phi$,
the scattering angles in the c.m.\ frame. However, as mentioned above, it
suffices to choose $\phi=0$. 

The remaining integrations over
$t$ and $t'$ in eqs.~(\ref{dij}) and (\ref{dtildeij}) are evaluated
numerically.  Details are presented 
in~\cite{Are07}. However, I would like to mention that in contrast to
eq.~(A.36) of ref.~\cite{Are07}, I found it more advantagous to use for
the numerical evaluation a
transformation of the integration variable $y=-\ln(1-t)$
resulting in 
\beqa
\int_0^1\frac{dt}{1-t}\,e^{-i\eta\ln{(1-t)}}(g(1)-g(t))&=&
\int_0^\infty dy\,e^{i\eta y}(g(1)-g(1-e^{-y}))\,.
\eeqa

 As mentioned in the appendix of ref.~\cite{Are07}, it is useful to
separate the contributions of the real and imaginary part of the radial
integral $I_{SS}$ for $\eta>0$ according to
\beqa
d^{[2]}_{ij}(\eta,\theta)&=&N(\eta)\Big({\cal R}^{SS}_{ij}(\eta,\theta)
+{\cal I}^{SS}_{ij}(\eta,\theta)\Big)\,,\label{dij+eta}
\eeqa
where ${\cal R/I}^{SS}_{ij}$ - from now on called reduced amplitudes -
refer to the contributions of $\,Re I_{SS}$
and $\,Im I_{SS}$, respectively (see eq.~(\ref{IntSS})), to $\widetilde{d}^{\,[2]}_{ij}(t,
\eta,\theta)$ in eq.~(\ref{dtildeij}). With respect to the symmetry
under sign change of $\eta$, one finds easily the properties
\beqa
{\cal R}^{SS}_{ij}(\theta,-\eta)&=&({\cal
  R}^{SS}_{ij}(\eta,\theta))^*\,,\\
{\cal I}^{SS}_{ij}(\theta,-\eta)&=&-({\cal
  I}^{SS}_{ij}(\eta,\theta))^*\,,
\eeqa
from which follows
\beqa
d^{[2]}_{ij}(\theta,-\eta)&=&N(-\eta)\Big({\cal R}^{SS}_{ij}(\eta,\theta)
-{\cal I}^{SS}_{ij}(\eta,\theta)\Big) ^*\,,\label{dij-eta}
\eeqa
Thus it suffices to evaluate the reduced amplitudes ${\cal
  R/I}^{SS}_{ij}$ for $\eta>0$ from which one can determine
$d^{[2]}_{ij}(\theta,\pm\eta)$ applying eqs.~(\ref{dij+eta}) and
(\ref{dij-eta}). On the other hand, given
$d^{[2]}_{ij}(\theta,\pm\eta)$ the reduced amplitudes are obtained
from 
\beqa
{\cal
  R}^{SS}_{ij}(\eta,\theta)&=&\frac{1}{2}\Big(\frac{d^{[2]}_{ij}(\eta,\theta)}{N(\eta)}+
\frac{d^{[2]}_{ij}(\theta,-\eta)^*}{N(-\eta)}\Big)
=\frac
{1}{2N(\eta)}\Big(d^{[2]}_{ij}(\eta,\theta)
+e^{-2\pi\eta}\,d^{[2]}_{ij}(\theta,-\eta)^* \Big)\label{RSS}
\,,\\
{\cal
  I}^{SS}_{ij}(\eta,\theta)&=&\frac{1}{2}\Big(\frac{d^{[2]}_{ij}(\eta,\theta)}{N(\eta)}-
\frac{d^{[2]}_{ij}(\theta,-\eta)^*}{N(-\eta)}\Big) =\frac
{1}{2N(\eta)}\Big(d^{[2]}_{ij}(\eta,\theta)
-e^{-2\pi\eta}\,d^{[2]}_{ij}(\theta,-\eta)^* \Big) \label{ISS}
\,.
\eeqa
These relations are useful for the comparison with the partial wave
approach. In addition, they show that  for large positive $\eta$ the
second terms in eqs.~(\ref{RSS}) and (\ref{ISS})  are strongly
suppressed by the factor $e^{-2\pi\eta}$. Consequently, one finds that 
${\cal R}^{SS}_{ij}(\eta,\theta)\approx {\cal
  I}^{SS}_{ij}(\eta,\theta)$ for positive $\eta \gg 1$. This feature poses a
serious problem for the numerical evaluation of
$d^{[2]}_{ij}(\theta,-\eta)$ for $\eta \gg 1$  because it is proportional to
the difference of the reduced amplitudes and multiplied with a huge
number ($N(-\eta)$) according to eq.~(\ref{dij-eta}). Thus with
increasing absolute 
value of $\eta$ more and more significant digits are lost in the
difference for negative $\eta$. An example is presented later.

\subsubsection{Spin-orbit interaction}
Following the analogous steps for the spin-orbit interaction, one
finds
\beq
\mathbf b(\theta,\phi)=i\,b_0(\eta,\theta)\frac{\mathbf p^{\,\prime}\times \mathbf p}
{|\mathbf p^{\,\prime}\times \mathbf p\,|}
\eeq
where
\beqa
b_0(\eta,\theta)&=&c^{LS}\,\sin\theta\,N(\eta)
\Big[\widetilde b_0(\theta,1,\eta)
+i\eta\int_0^1\frac{dt}{1-t}\,e^{-i\eta\ln{(1-t)}}
\Big(\widetilde b_0(\theta,1,\eta) -e^{i\eta\ln{t}}t\,\widetilde
b_0(\theta,t,\eta) \Big) \,,\\
\widetilde b_0(\theta,t,\eta)&=&\frac{I_{LS}(c(t,1))}{g(t,1)^2}
+i\eta\int_0^1\frac{dt'}{1-t'}\,e^{-i\eta\ln{(1-t')}}
\Big(\frac{I_{LS}(c(t,1))}{g(t,1)^2}
-e^{i\eta\ln{t'}}\frac{t' I_{LS}(c(t,t'))}{g(t,t')^2} \Big) \,,
\eeqa
with $g(t,t')$ in eq.~(\ref{gtts}) and the radial integral
 \beqa
I_{LS}(c)&=& \int_0^\infty\frac{dx}{x}\,e^{icx}\,j_1(x)=1
-\frac{c}{2}\,\ln\Big|\frac{c+1}{c-1}\Big|+i\frac{\pi c}{2}\,\Theta(1-c)\,.
\eeqa
The numerical evaluation is analogous to the one for the hyperfine
interaction. For $\eta=0$ one finds 
$b_0(0,\theta)=c^{LS}\,\sin\theta/(4\sin^2\theta/2)$ and thus
\beq
\mathbf b^{PW}(\theta,\phi)=\frac{i}{2}\,\cot{(\theta/2)}\,c^{LS}\,
\frac{\mathbf p^{\,\prime}\times \mathbf p}{|\mathbf p^{\,\prime}\times \mathbf p\,|}
\eeq
in agreement with eq.~(\ref{b-pw}).

Also in this case I introduce reduced amplitudes by separating the
contributions of the real and imaginary 
parts of the radial integral $I_{LS}$ acording to
\beqa
b_{0}(\eta,\theta)&=&N(\eta)\Big({\cal R}^{LS}(\eta,\theta)
+{\cal I}^{LS}(\eta,\theta)\Big)\,,
\eeqa
such that
\beqa
b_{0}(\theta,-\eta)&=&N(-\eta)\Big({\cal R}^{LS}(\eta,\theta)
-{\cal I}^{LS}(\eta,\theta)\Big)^*\,,\label{b0-eta}
\eeqa
and
\beqa
  {\cal R}^{LS}(\eta,\theta)&=&\frac{1}{2 N(\eta)}\Big(b_0(\eta,\theta)+
e^{-2\pi\eta}\,b_0(\theta,-\eta)^*\Big)\,,\\
 {\cal  I}^{LS}(\eta,\theta)&=&\frac{1}{2 N(\eta)}\Big(b_0(\eta,\theta)-
e^{-2\pi\eta}\,b_0(\theta,-\eta)^*\Big)
\,.
\eeqa
Again numerical problems arise for negative $\eta$ with
$|\eta|\gg 1$ as for the hyperfine interaction outlined above.

\subsection{Partial wave expansion}

The expansion of the Coulomb wave function into partial waves reads 
\beq
\psi^{(+)}_{\mathbf p}(\mathbf r\,)=\frac{4\pi}{pr}\sum_{l,m}\,i^le^{i\bar\sigma_l}\,
F_l(\eta,pr) Y_{lm}^*(\hat r)\,Y_{lm}(\hat p)
\eeq
where the radial function $F_l$ is given in terms of the confluent
hypergeometric function $_1F_1(a,b,z)$
\beqa
F_l(\eta,\rho)&=&C_l (\eta)\,e^{i\rho}\,\rho^{l+1} \,_1F_1(l+1+i\eta,2l+2,-2i\rho)\,,
\eeqa
and constants depending on the Sommerfeld parameter $\eta$
\beqa
C_l(\eta)&=&\frac{2^l}{(2l+1)!}e^{-\frac{\pi}{2}\eta}|\Gamma(l+1+i\eta)|\nonumber\\
&=&\frac{C_0(\eta)}{(2l+1)!!}\,D_l(\eta^2)\,,\\
C_0(\eta)&=&e^{-\frac{\pi}{2}\eta}|\Gamma(1+i\eta)|=e^{-\frac{\pi}{2}\eta}
\sqrt{\frac{\pi \eta}{\sinh(\pi\eta)}}\,,\\
D_l(\eta^2)&=&\prod_{n=1}^l\sqrt{1+\Big(\frac{\eta}{n}\Big)^2}\,.
\eeqa
In the above expression, I have separated as in
eq.~(\ref{c-scatt-wave}) the $l=0$ phase $\sigma_0=\sigma_C$ for
convenience, coinciding with the Coulomb phase $\sigma_C$ 
 given in eq.~(\ref{coulomb-phase}). The remaining partial wave
 phase is given by
\beq
\bar\sigma_l=\sigma_l-\sigma_0\,,
\eeq
where for $l>0$
\beq
e^{i \bar\sigma_l}=\frac{l+i\eta}{|l+i\eta|}\cdots \frac{1+i\eta}{|1+i\eta|}\,.
\eeq
Evaluation of the various contributions
to the scattering matrix as listed in eqs.~(\ref{ls_dwba}) and
(\ref{t_dwba_2}) leads to the following expressions which are 
still operators in spin space
\beqa
\mathbf b_{l/h}^{DW}\cdot\boldsymbol\sigma_{l/h}&=&\sum_{l=1}^\infty G_{LS_{l/h}}^l 
\mathbf\Omega_{ll}(\hat p',\hat p) \cdot\boldsymbol\sigma_{l/h}\,,\label{T-LS}\\
\sum_{ij}\sigma_{l,i} d^{[2]}_{ij}\sigma_{h,j}
\,&=&\sum_{l'=0}^\infty \sum_{l=0}^\infty
G_{SS,2}^{l'l} 
\Big[\Sigma^{[2]}(\boldsymbol\sigma_l,\boldsymbol\sigma_h)\times
\Omega_{l'l}^{[2]}\Big]^{[0]}(\hat p',\hat p)\,,\label{T-SS2}
\eeqa
where I have introduced for convenience in the notation of Fano and
Racah~\cite{FaR59} for irreducible spherical tensors 
\beqa
\Omega_{l'l}^{[K]}(\hat p',\hat p)&=&\Big[Y^{[l']}(\hat p')\times
Y^{[l]}(\hat p)]\Big]^{[K]}\,,\\
\Sigma^{[2]}(\boldsymbol\sigma_l,\boldsymbol\sigma_h)&=&\Big[\sigma_{l}^{[1]}\times
\sigma_{h}^{[1]}\Big]^{[2]} \,.
\eeqa
The coefficients are given in terms of the radial matrix elements
$R_{l'l}$ (note the meaning of the``hat symbol'': $\hat l=\sqrt{2l+1}$)
\beqa
G_{LS_{l/h}}^l (\eta)&=& (-)^{l+1}\frac{4\pi}{\sqrt{3}} c_{l/h}^{LS}\hat
l\sqrt{l(l+1)}e^{2i\bar\sigma_l}\, 
R_{ll}\,.\\
%G_{SS,0}&=&-\frac{2}{\sqrt{3}}\,(4\pi)^2\,c^{SS}c_0^2\,,\\
G_{SS,2}^{l'l}(\eta)&=& i^{l-l'}16\,\pi \sqrt{6}\,c^{SS}\,
\hat {l'} \hat l \,e^{i(\bar\sigma_{l'} +\bar\sigma_l)}\,
\left(\begin{matrix}
l^{\prime}&l& 2 \cr 0 &0&0 \cr
\end{matrix}\right)
R_{l'l}\,.
\label{gss2}
\eeqa
with
\beqa
R_{l'l}&=&\frac{4}{p^2}\int_0^\infty
\frac{dr}{r^3}\,F_{l'}(\eta,pr)F_l(\eta,pr)
=4\int_0^\infty
\frac{d\rho}{\rho^3}\,F_{l'}(\eta,\rho)F_l(\eta,\rho)\,.
\eeqa
Besides the radial integral $R_{ll}$, only $R_{l'l}=R_{ll'}$ for
$|l-l'|=2$ are needed in 
view of the selection rule of the 3j-symbol in eq.~(\ref{gss2}).
This radial integral is well known in Coulomb excitation (see
e.g.~\cite{BiB65}). 
The explicit form of the radial integral is also given
in~\cite{MiS08}. For $l'=l$ and $l>0$ one has
\beqa
R_{ll}&=&\frac{2}{ l(l+1)}\,\Big (1+\frac{f_l(\eta)}{2l+1}\Big)
\label{rll}
\eeqa
with
\beqa
f_l(\eta)&=&e^{-\pi\eta}\,\frac {\pi\eta}{\sinh{(\pi\eta)}}-1
-2\eta^2\sum_{k=1}^l\frac{1}{k^2+\eta^2}\,.
\eeqa
One should note that $f_l$ vanishes for $\eta=0$. 
For $|l'-l|=2$ one has
\beq
R_{l,l+2}=\frac{2}{3|l+1+i\eta||l+2+i\eta|}\,. \label{rlls}
\eeq
\subsubsection{The hyperfine contribution}
The tensor amplitude $d^{[2]}_{ij}$ of the hyperfine interaction is
obtained by separating the spin dependence
in eq.~(\ref{T-SS2}). This means, one has to evaluate
\beqa
d^{[2]}_{ij}&=&\frac{\partial^2}{\partial \sigma_{l,i}\partial
  \sigma_{h,j}}\sum_{ij}\sigma_{l,i} d^{[2]}_{ij}\sigma_{h,j}
\nonumber\\
&=&\sum_{l'=0}^\infty \sum_{l=0}^\infty
G_{SS,2}^{l'l} \frac{\partial^2}{\partial \sigma_{l,i}\partial
  \sigma_{h,j}}\Big[\Sigma^{[2]}(\boldsymbol\sigma_l,\boldsymbol\sigma_h)\times
\Omega_{l'l}^{[2]}\Big]^{[0]}(\hat p',\hat p)\,.
\eeqa
It suffices to consider $d^{[2],0}_{ij}$ for
the special case, for which the scattering plane 
coincides with the $x$-$z$-plane, i.e.\ $\hat
p^{\,\prime}=(\sin\theta,0,\cos\theta)$.  First one notes that then
\beqa
\Omega_{l'l,m}^{[2]}(\hat p',\hat p)&=&\frac{\sqrt{5}}{4\pi}(-)^{l+l}\hat
  l^{\,\prime}\hat l\sum_m\sqrt{\frac{(l-m)!}{(l+m)!}}
\left(\begin{matrix}
l'&l&2\cr
m&0&-m\cr
\end{matrix}\right)
P_l^m(\cos\theta)\,,
\eeqa
where $P_l^m$ denotes the associated Legendre function,
and thus
\beq
\frac{\partial^2}{\partial \sigma_{l,i}\partial
  \sigma_{h,j}}\Big[\Sigma^{[2]}(\boldsymbol\sigma_l,\boldsymbol\sigma_h)\times
\Omega_{l'l}^{[2]}\Big]^{[0]}=\frac{\sqrt{5}}{4\pi}(-)^{l+l}\hat
  l^{\,\prime}\hat l\sqrt{\frac{(l-m)!}{(l+m)!}}
\left(\begin{matrix}
l'&l&2\cr
m&0&-m\cr
\end{matrix}\right)
P_l^m(\cos\theta) \sigma_{ij}^m\,,
\eeq
where
\beqa
\sigma_{ij}^m&=&\frac{\partial^2}{\partial \sigma_{l,i}\partial
  \sigma_{h,j}}
\Sigma^{[2]}_m(\boldsymbol\sigma_l,\boldsymbol\sigma_h)\,.
\eeqa
With these expressions one finds for the tensor part of the hyperfine
contribution 
\beqa
d^{[2],0}_{ij}=\frac{\sqrt{3}}{2\sqrt{2}}\,c^{SS}\,
\sum_l ^\infty &&\Big[ S_{l}^0P_l(\cos{\theta})\sigma_{ij}^0
+S_{l}^1\sqrt{\frac{(l-1)!}{(l+1)!}}P_l^1(\cos{\theta}) \Big
(\sigma_{ij}^1-\sigma_{ij}^{-1}\Big) \nonumber\\&&
+S_{l}^2\sqrt{\frac{(l-2)!}{(l+2)!}}P_l^2(\cos{\theta}) \Big
(\sigma_{ij}^2+\sigma_{ij}^{-2}\Big) \Big]\,,
\eeqa
where for $m=0,1,2$
\beqa
S_{l}^m&=&(-i)^{l}\hat{l}^{\,2}
e^{i\bar\sigma_l}\sum_{k=|l-2|}^{l+2}i^k\hat k^{\,2} e^{i\bar\sigma_k}
\left(\begin{matrix}
l &k&2\cr
0&0&0\cr
\end{matrix}\right)
\left(\begin{matrix}
l &k&2\cr
-m&0&m\cr
\end{matrix}\right) R_{lk}\,.
\eeqa
Explicitly, with
\beq
\sigma_{11/22}^m= -\frac{ \delta_{m0}}{\sqrt{6}}\pm \frac{\delta_{|m|2}}{2}\,,\quad 
\sigma_{33}^m= \delta_{m0}\sqrt{\frac{2}{3}}\,,\quad 
\sigma_{12}^m= i\frac{m}{2}\delta_{|m|2}\,,\quad 
\sigma_{13}^m= -\frac{m}{2}\delta_{|m|1}\,,\quad 
\sigma_{23}^m= -\frac{i}{2}\delta_{|m|1}\,,
\eeq
 one obtains for the nonvanishing components
\beqa
d^{[2],0}_{33}&=&c^{SS}\,\sum_{l=0} ^\infty 
S_{l}^{33}\,P_l(\cos{\theta})\,,\\
d^{[2],0}_{11/22}&=&\pm\,c^{SS}\,\sum_{l=2} ^\infty 
S_{l}^{11}\,P_l^2(\cos{\theta})-\frac{1}{2}d^{[2],0}_{33}\,,\\ 
d^{[2],0}_{13}&=&c^{SS}\,\sum_{l=1} ^\infty 
S_{l}^{13}\,P_l^1(\cos{\theta})\,,
\eeqa
where I have introduced for convenience
\beqa
S_l^{33}&=&\frac{1}{2}\,S_{l}^0\,,\quad
S_l^{11}=\frac{1}{2}\,\sqrt{\frac{3(l-2)!}{2(l+2)!}}\,S_{l}^2\,,\quad
S_l^{13}=-\frac{1}{2}\,\sqrt{\frac{3(l-1)!}{2(l+1)!}}\,S_{l}^1\,.
\eeqa

It is useful to separate the $\eta$-independent contributions,
constituting the plane wave approximation. One 
finds explicitly the following detailed expressions
\beqa
d^{[2],0}_{33}(\eta)&=&c^{SS}\,\Big (\sin^2(\theta/2)-\frac{1}{3}\,
+\sum_{l=0} ^\infty 
\widetilde S_{l}^{33}(\eta)\,P_l(\cos{\theta}) \Big)\,,\\
d^{[2],0}_{11/22}(\eta)&=&\pm\,c^{SS}\,\Big (\frac{1}{2}\,\cos^2(\theta/2)
+\sum_{l=2} ^\infty 
\widetilde S_{l}^{11}(\eta)\,P_l^2(\cos{\theta}) \Big)-\frac{1}{2}d^{[2],0}_{33}\,,\\ 
d^{[2],0}_{13}(\eta)&=&c^{SS}\,\Big(-\frac{1}{2}\sin(\theta)+
\sum_{l=1} ^\infty 
\widetilde S_{l}^{13}(\eta)\,P_l^1(\cos{\theta}) \Big)\,,
\eeqa
where the coefficients $\widetilde S_{l}^{ij}(\eta)$ vanish for
$\eta=0$. In detail one finds for $i=j=3$ and $l=0,1$
\beqa
\widetilde S_{0}^ {33}(\eta)&=&\frac{ i\eta}{3}\,\frac{3-i\eta}{(1-i\eta)(2-i\eta)} \,,\\
\widetilde S_{1}^ {33}(\eta)&=&-\frac{3}{5}\,
\Big[\frac{1+i\eta}{1-i\eta}\frac{f_1(\eta)}{3}
-\frac{i\eta(5-i\eta)}{6(2-i\eta)(3-i\eta)} 
-\frac{2i\eta}{1-i\eta}\Big(1-\frac{1 }{(2-i\eta)(3-i\eta)}
\Big) \Big]\,,
\eeqa
and for $l>1$
\beqa
\widetilde S_{l}^ {33}(\eta)&=&-e^{2i\bar\sigma_l}
\Big[
\frac{i\eta}{2}\Big(b_l(\eta)-b_{l+2}(\eta)^*\Big)
+\frac{f_l(\eta)}{(2l-1)(2l+3)}
\Big]\,.
\eeqa
Here I have introduced for convenience
\beqa
b_l(\eta)&=&\frac{2l-1 +i\eta}{(2l-1)(l-1)l(l-1 +i\eta)(l +i\eta)}\,.
\eeqa
One should note that the coefficients $\widetilde S_{l}^ {33}$ behave
as $1/l^2$ for $l\to\infty$. 

For $i=j=1$ one obtains (note $l>1$)
\beqa
\widetilde  S_{l}^ {11} (\eta)&=&
\frac{(2l+1)( 1-e^{2i\bar\sigma_{l}})}{2(l-1)l(l+1)(l+2)}
+e^{2i\bar\sigma_{l}}\Big(\frac{i\eta}{4}(b_l(\eta)-b_{l+2}(\eta)^*)
-\frac{3f_l(\eta)}{2l(l+1)(2l-1)(2l+3)}\Big)\,,
\eeqa
The coefficient $ \widetilde S_{l}^ {11}$ behaves as $l^{-3}$ for
$l\to\infty$. 

Finally, for 
$ \widetilde S_{l}^ {13}$ one obtains for $l=1$
\beqa
\widetilde S_{1}^ {13}(\eta) &=&\frac{3}{20}\,
\Big[\frac{1+i\eta}{1-i\eta}\,f_1(\eta)
+\frac{5}{3} i\eta b_3(\eta)^*
+\frac{2i\eta}{1-i\eta}\Big(3+\frac{2 }{(2-i\eta)(3-i\eta)}
\Big) \Big]
\eeqa
and for $l>1$
\beqa
\widetilde S_{l}^ {13} (\eta)&=&\frac{1}{2}e^{2i\bar\sigma_l}
\Big(\frac{i\eta}{4}(b_l(\eta)+b_{l+2}(\eta)^*)
+\frac{3f_l(\eta)}{l(l+1)(2l-1)(2l+3)}\Big)\,
\eeqa
The coefficient $\widetilde S_{l}^ {13} $ behaves like $l^{-3}$ for $l\to\infty$
and vanishes for $\eta=0$.  

\begin{figure}[ht]
\includegraphics[width=.7\columnwidth]{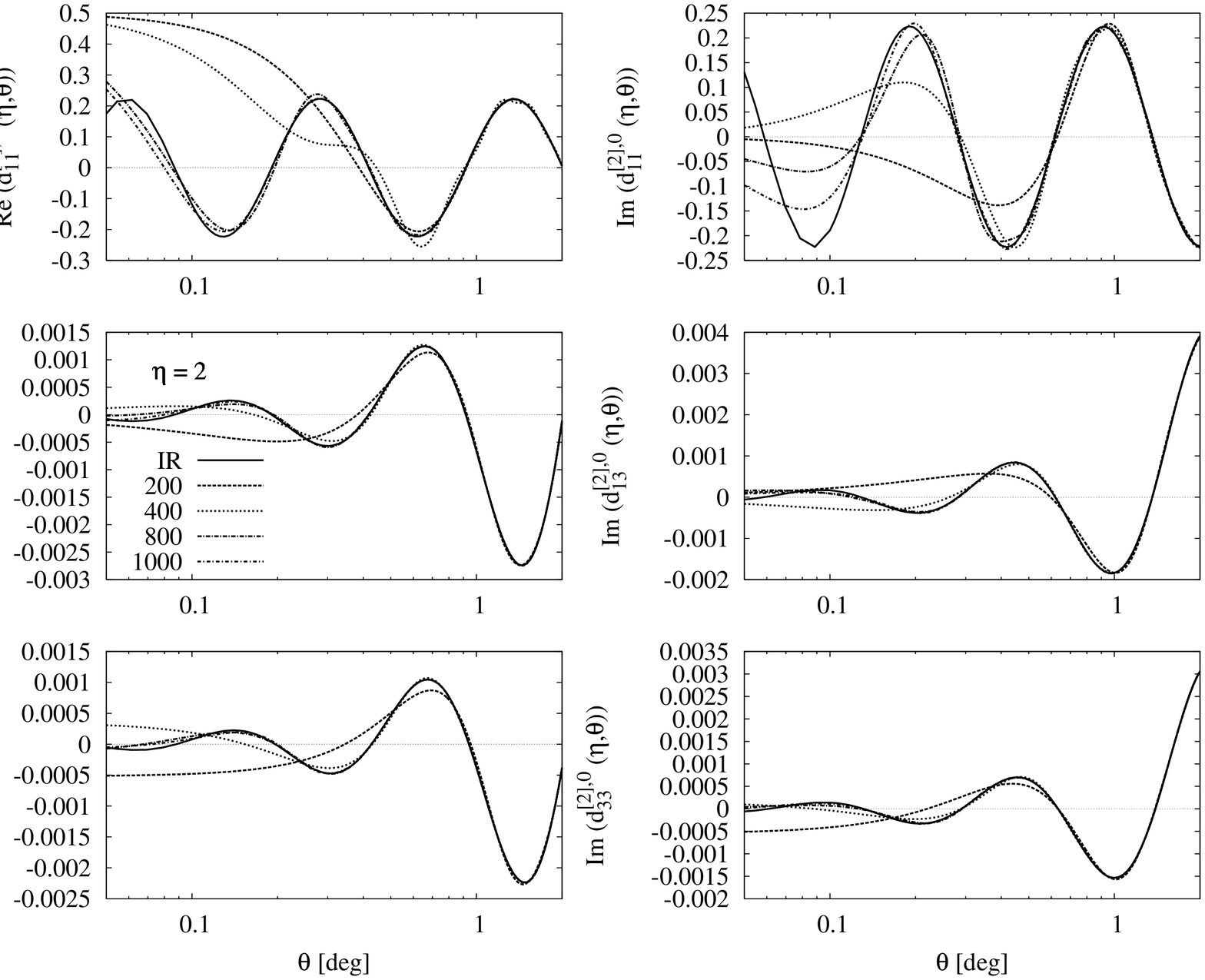}
\caption{Convergence of the partial wave expansion (PWE) of the hyperfine
  amplitudes $d_{ij}^{[2],0}(\eta,\theta)$  for various $l_{max}$ as indicated in the legend
for $\eta=2$ for $(ij)=(11)$ (upper panels), $(ij)=(13)$ (middle
panels), and $(ij)=(33)$ (lower panels). For comparison the result of the integral
representation (IR, solid curves) is also shown.}
\label{Fig_convergence_dij} 
\end{figure}

The convergence of the partial wave series is quite good in general as is
demonstrated in Fig.~\ref{Fig_convergence_dij} for $\eta=2$. Only
$d_{11}^{[2],0}(\eta,\theta)$ shows a slower convergence at very small
angles. 

\subsubsection{The spin-orbit contribution}
\begin{figure}[ht]
\includegraphics[width=.7\columnwidth]{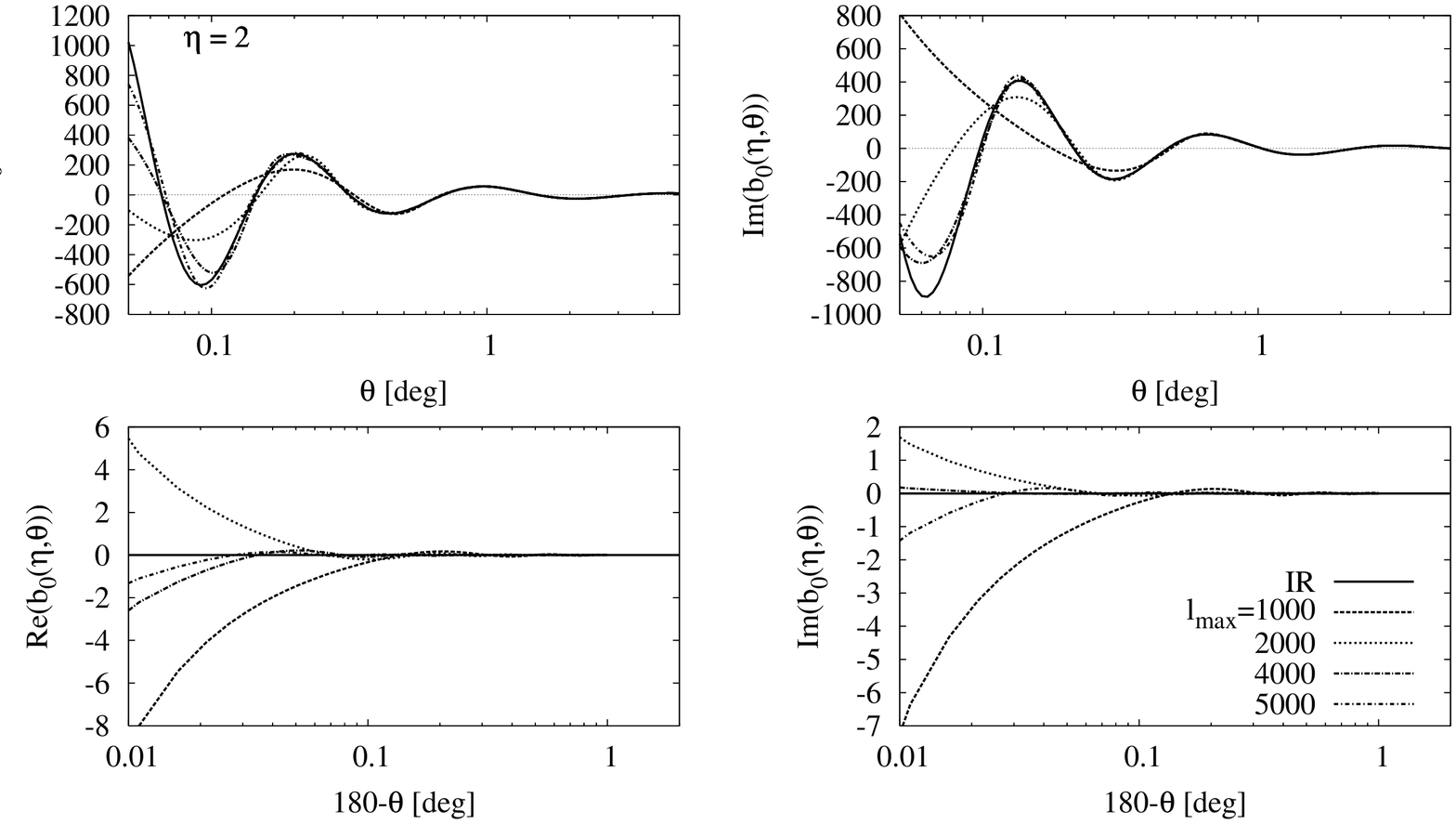}
\caption{Convergence of the partial wave expansion (PWE) of the
  spin-orbit amplitude $b^0(\eta,\theta)$   for various $l_{max}$ as indicated in the legend 
for $\eta=2$ near $\theta=0^\circ$ (upper panels) and near $180^\circ$
(lower panels).  For comparison the result of the integral
representation (IR, solid curves) is also shown.}  
\label{Fig_convergence_b0} 
\end{figure}

Acording to eq.~(\ref{T-LS}) the spin-orbit strength is given by
\beqa
\mathbf b&=&\sum_{l=1}^\infty G_{LS}^l
\mathbf \Omega_{ll}(\hat p',\hat p)\,.
\eeqa
For the chosen reference frame, i.e.\ $\hat p=(0,0,1)$
and $\hat
p^{\,\prime}=(\cos\phi\sin\theta,\sin\phi\sin\theta,\cos\theta)$, one  
obtains 
\beqa
\Omega_{ll,x/y}(\hat p',\hat p)&=&i\frac{\hat
  l^{\,2}}{4\pi}\sqrt{\frac{6}{l(l+1)}} 
\left(\begin{matrix}
l &l&1\cr
1&0&-1\cr
\end{matrix}\right)
P_l^1(\cos\theta)
\left\{\begin{array}{r}
-\sin\phi\cr
\cos\phi\cr
\end{array}\right\}\,,\\
\Omega_{l'l,z}(\hat p',\hat p)&=&0\,,
\eeqa
and thus the spin-orbit vector $\mathbf b$ has the form
\beqa
\mathbf b =i\,b_0\,\frac{\mathbf p^{\,\prime}\times\mathbf p}
{|\mathbf p^{\,\prime}\times\mathbf p\,|}
\eeqa
with
\beqa
b_0(\eta,\theta)&=&-\frac{1}{2}{c^{LS}}\,\sum_{l=1}^\infty 
\beta_l(\eta) P^1_l(\cos{\theta})\,,
\eeqa
where
\beqa
\beta_l(\eta) &=&\frac{1}{2}\,\hat l^2\,e^{2i\bar\sigma_l}R_{ll}
=\frac{e^{2i\bar\sigma_l}}{l(l+1)}(2l+1+f_l(\eta))\,.
\eeqa
This form is not well suited for a numerical evaluation, because even
for $\eta=0$ the sum extends up to infinity. Therefore,  it is more
advantageous to separate the $\eta $-independent part writing
\beqa
\beta_l(\eta) &=&\frac{2l+1}{l(l+1)} + \beta_l^\eta 
\eeqa
with
\beq
\beta_l^\eta =\frac{e^{2i\bar\sigma_l}}{l(l+1)}
\Big((2l+1)(e^{2i\bar\sigma_l}-1)+f_l(\eta)\Big)\,.
\eeq
The coefficient $\beta_l^\eta$ vanishes for $\eta=0$. 
For the $\eta$ independent part one can rearrange the sum by using 
\beq
P^1_l(x)=\frac{l(l+1)}{\hat l^2\sqrt{1-x^2}}\Big(P_{l+1}(x)-P_{l-1}(x)\Big)\,,\label{Pl1}
\eeq
and one finds 
\beqa
\sum_{l=1}^\infty \frac{2l+1}{l(l+1)}  P^1_l(x)&=&
\frac{1}{\sqrt{1-x^2}}\Big(\sum_{l=2}^\infty P_l(x) -\sum_{l=0}^\infty
P_l(x) \Big)\nonumber\\ 
&=&-\frac{1}{\sqrt{1-x^2}}(P_0(x)+P_1(x))=-\cot(\theta/2)\,.
\eeqa
Then one obtains for $b_0(\eta,\theta)$
\beqa
b_0(\eta,\theta)&=&\frac{1}{2}\,c^{LS}
\Big(\cot(\theta/2)-\sum_{l=1}^\infty 
\beta_l^\eta P^1_l(\cos{\theta})\Big)\,,\label{sumbLS_a}
\eeqa
yielding $b_0(0,\theta)$
in accordance with eq.~(\ref{b-pw}).
One can evaluate directly this expression or rearrange also the
remaining sum yielding
\beqa
b_0(\eta,\theta)&=&\frac{1}{2}\,c^{LS}
\Big(\cot(\theta/2)-\frac{1}{\sin{\theta}}
\sum_{l=0}^{\infty} e_l(\eta)\,P_l
(\cos{\theta}) \Big]\,, \label{sumbLS_b}
\eeqa
where for $l=0,1$
\beq
e_l(\eta)= \frac{(l+1)(l+2)}{2l+3}\beta_{l+1}^\eta\,,
\eeq
and for $l>1$
\beqa
e _l(\eta)&=&\frac{(l+1)(l+2)}{2l+3}\beta_{l+1}^\eta-\frac{(l-1)l}{2l-1}\beta_{l-1}^\eta
\nonumber\\
&=&2\,e^{2i\bar\sigma_l}\Big [i\eta\,\frac{2l+1}{(l+i\eta)(l+1-i\eta)}
-\frac{f_l(\eta)}{(2l-1)(2l+3)}\Big(2-i\eta\Big\{\frac{2l-1}{l+1-i\eta}
+\frac{2l+3}{l+i\eta}\Big\}\Big) \nonumber\\
&&-\eta^2\Big(\frac{1}{(2l+3)(l+1-i\eta)^2}+\frac{1}{(2l-1)(l+i\eta)^2}\Big)\Big]\,. 
\eeqa
For $l\to\infty$ the coefficients $\beta_l^\eta$ and $e _l$ behave as
$1/l$ resulting in a considerably slower convergence than for the
hyperfine amplitude. This is
demonstrated for the region of small angles and near $\theta=180^\circ$ in
Fig.~\ref{Fig_convergence_b0}  for $\eta=2$ for 
various $l_{max}$-values up to $l_{max}=5000$ with the expansion of
eq.~(\ref{sumbLS_b}). For comparison the 
result for the integral representation is also shown. The expansion of
eq.~(\ref{sumbLS_a}) gives for large absolute values of $\eta$ the
same result. However, for small $\eta$ it results in small
oscillations around the result of the other expansion at small
angles. 

\subsection{Comparison of the two methods}

\begin{figure}[ht]
\includegraphics[width=.7\columnwidth]{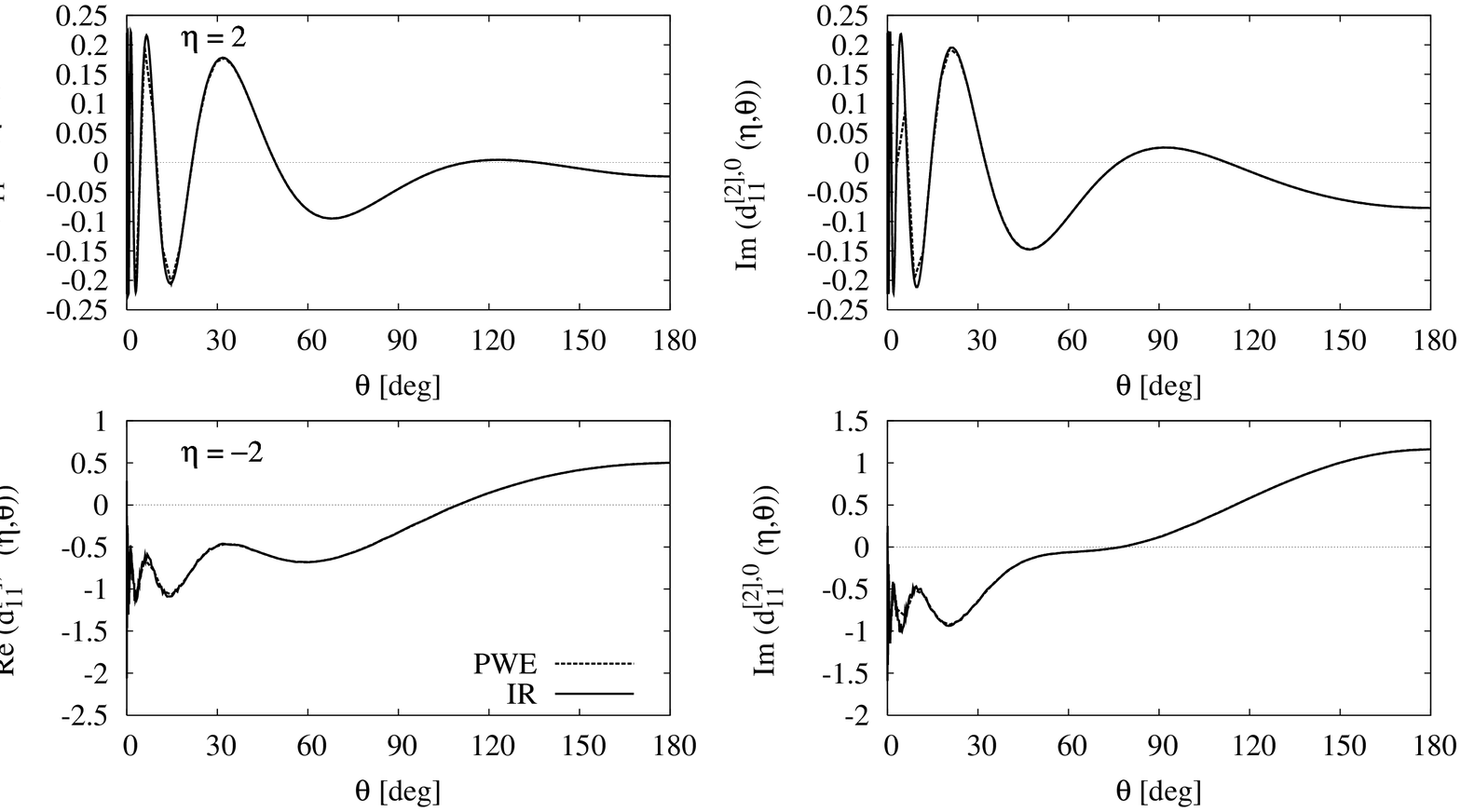}
\caption{Hyperfine amplitude $d^{[2],0}_{11}(\eta,\theta)$ 
for $\eta=\pm2$ for the integral
representation (IR) and the partial wave expansion (PWE).} 
\label{Fig_comparison_d11SS} 
\end{figure}

\begin{figure}[ht]
\includegraphics[width=.7\columnwidth]{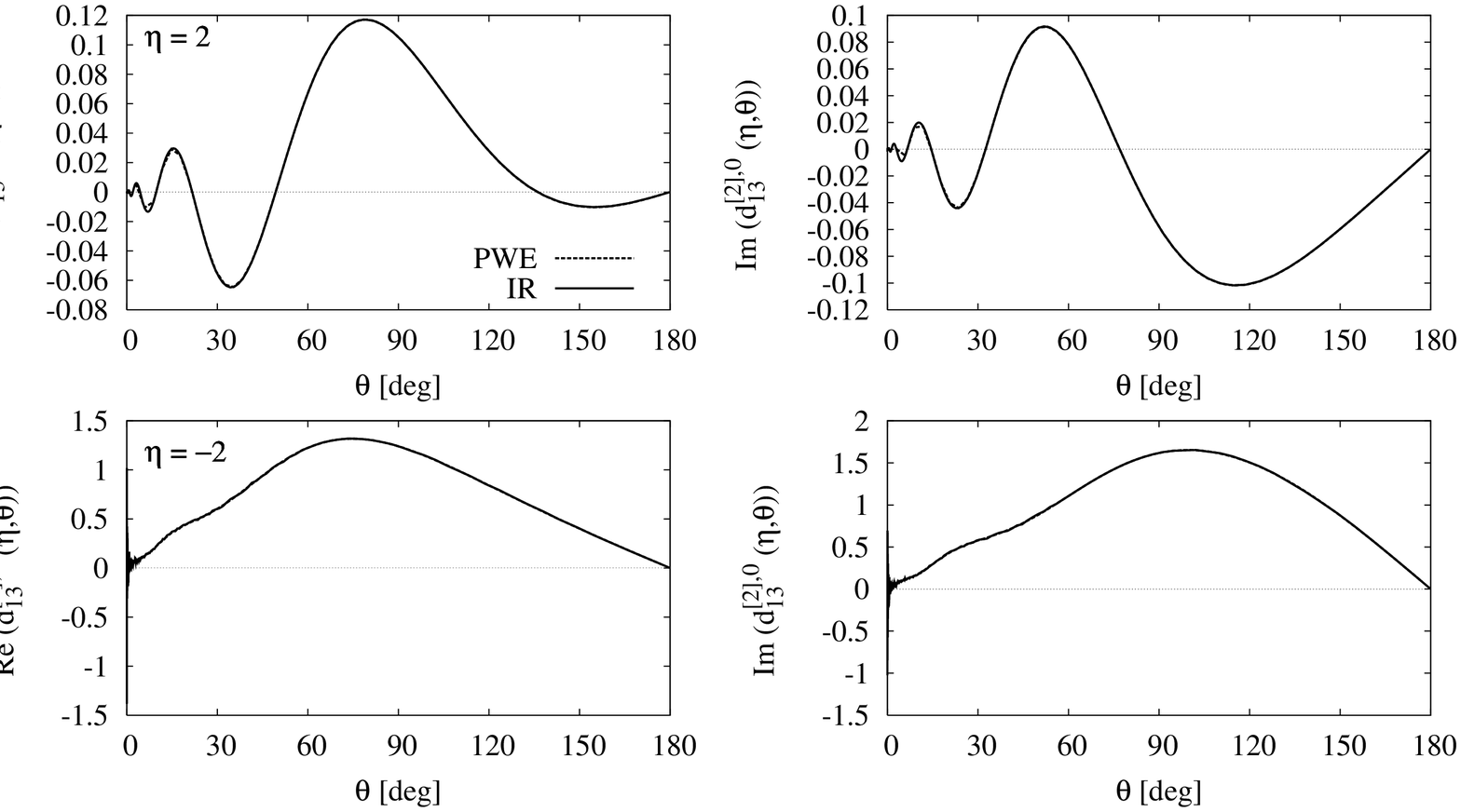}
\caption{Hyperfine amplitude $d^{[2],0}_{13}(\eta,\theta)$ 
for $\eta=\pm2$ for the integral
representation (IR) and the partial wave expansion (PWE).} 
\label{Fig_comparison_d13SS} 
\end{figure}

\begin{figure}[ht]
\includegraphics[width=.7\columnwidth]{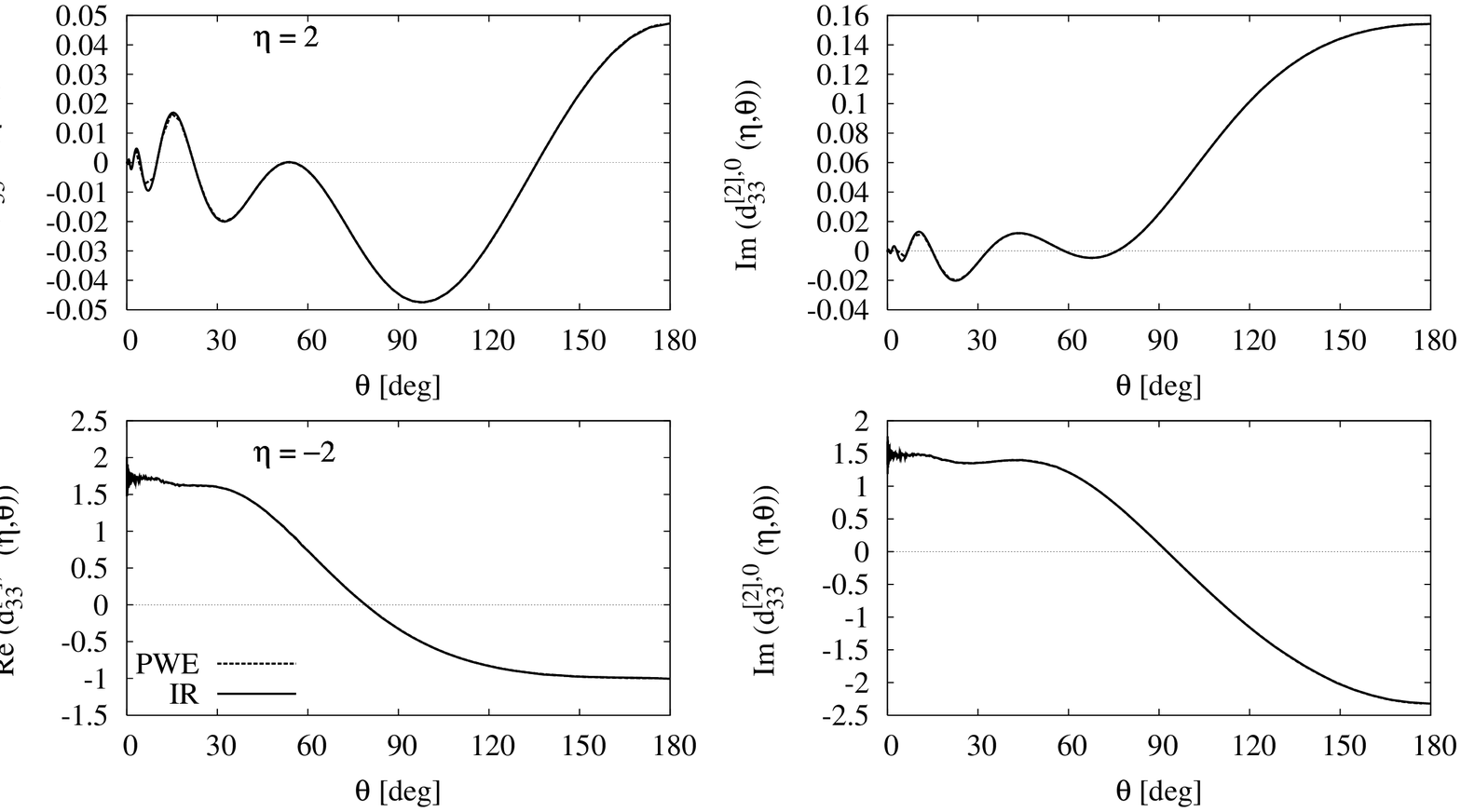}
\caption{Hyperfine amplitude $d^{[2],0}_{33}(\eta,\theta)$ 
for $\eta=\pm2$ for the integral
representation (IR) and the partial wave expansion (PWE).} 
\label{Fig_comparison_d33SS} 
\end{figure}

\begin{figure}[ht]
\includegraphics[width=.7\columnwidth]{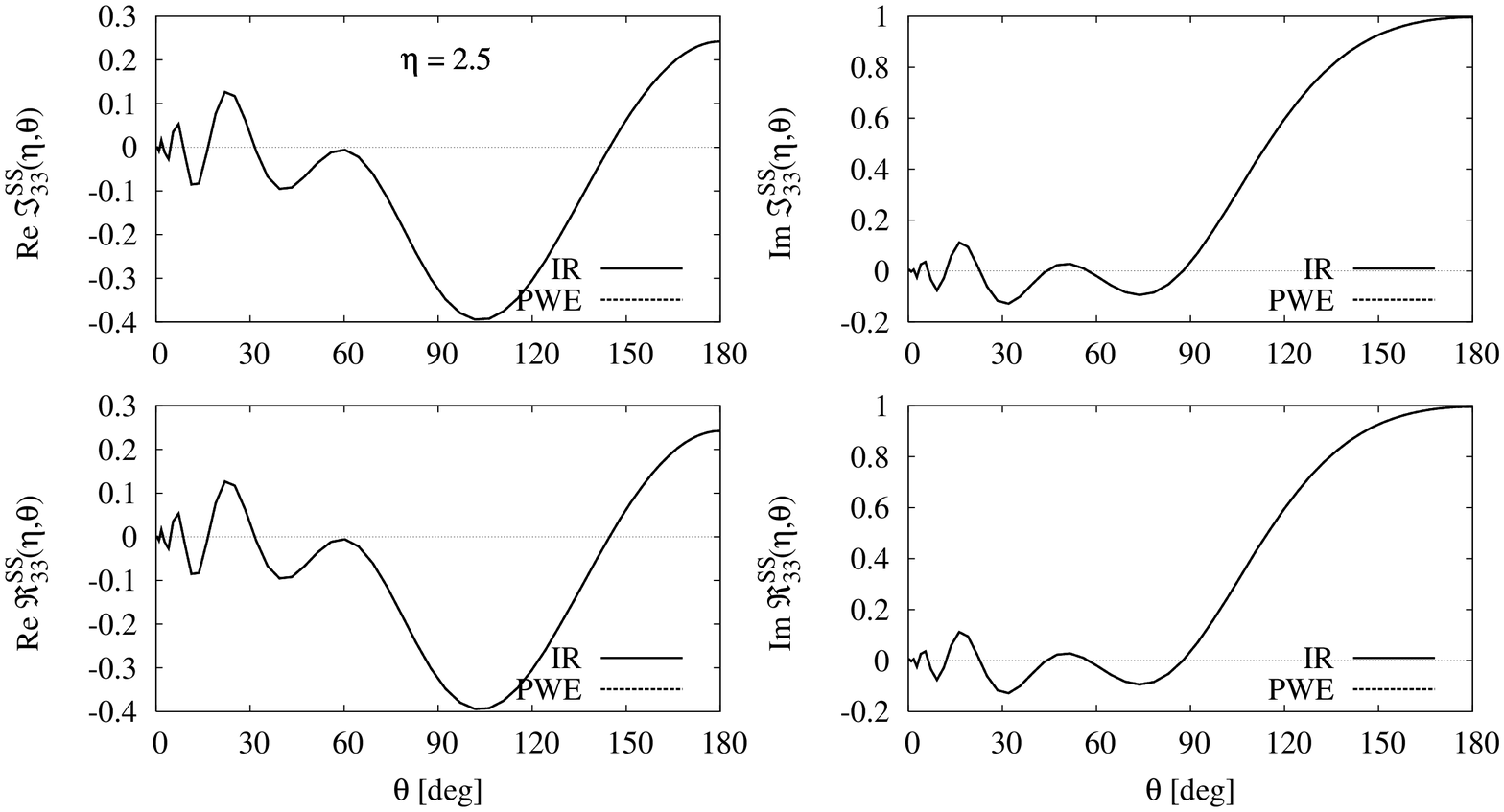}
\caption{Reduced amplitudes ${\cal  R}^{SS}_{33}(\eta,\theta)$  and
${\cal  I}^{SS}_{33}(\eta,\theta)$ for $\eta=2.5$ for the integral
representation (IR) and the partial wave expansion (PWE).} 
\label{Fig_comparison_SS_R_I} 
\end{figure}

\begin{figure}[ht]
\includegraphics[width=.7\columnwidth]{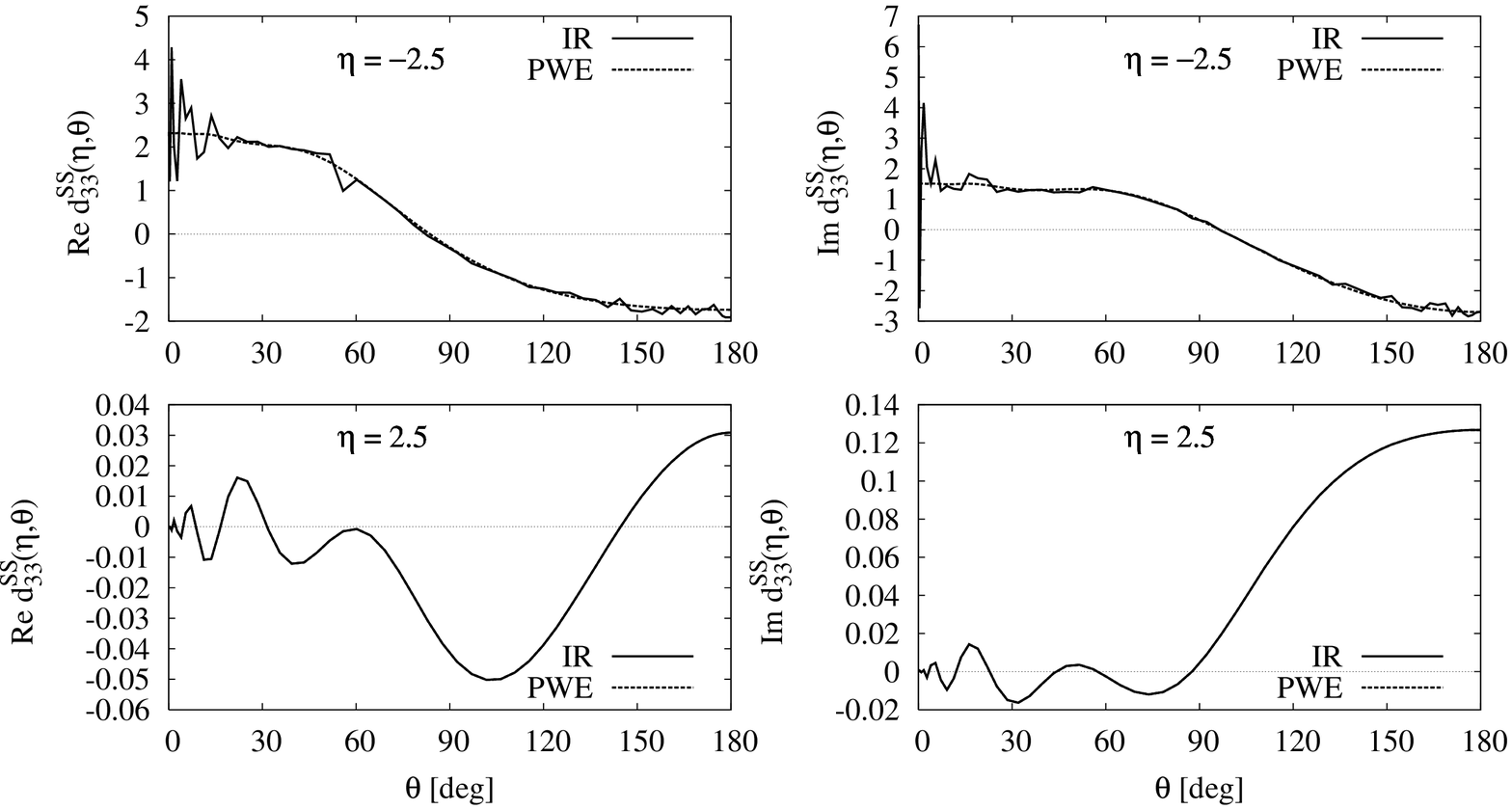}
\caption{Hyperfine amplitude $d_{33}^{[2],0}(\eta,\theta)$ 
for $\eta=\pm 2.5$ for the integral
representation (IR) and the partial wave expansion (PWE).} 
\label{Fig_comparison_d33} 
\end{figure}

For the hyperfine and spin-orbit interactions both methods give
identical results for the corresponding amplitudes as
is demonstrated for the hyperfine amplitudes in
Figs.~\ref{Fig_comparison_d11SS} through
\ref{Fig_comparison_d33SS} for $\eta=\pm 2$. The above mentioned numerical
problem which arises in the integral representation for higher
negative $\eta$ is illustrated in Fig.~\ref{Fig_comparison_SS_R_I} where, as an
example, the reduced amplitudes ${\cal  R}^{SS}_{33}(\eta,\theta)$
and ${\cal I}^{SS}_{33}(\eta,\theta)$ for $\eta=2.5$ are plotted. The two curves labeled
``IR'' and ``PWE'' are almost indistinguishable. This means also
complete agreement for the amplitudes for positive $\eta >0$ as shown in the
lower two panels of Fig.~\ref{Fig_comparison_d33} for the amplitude
$d_{33}^{[2]}(\eta,\theta)$. Furthermore, one notes that for this 
$\eta$-value the two reduced amplitudes in
Fig.~\ref{Fig_comparison_SS_R_I} are almost equal to a high 
degree of accuracy. Thus, 
this feature creates the numerical problem mentioned above for the integral
representation method, because 
for negative values the amplitudes are represented as differences of the reduced
amplitudes (see eqs.~(\ref{dij-eta}) and
(\ref{b0-eta})). Consequently, large cancellations occur which reduce
the numerical accuracy more and more with inreasing absolute values
for negative $\eta$. In fact, accuracy is lost for about $-\eta>
2$. This is demonstrated in the upper two panels of
Fig.~\ref{Fig_comparison_d33} for $\eta=-2.5$ where one readily notes
the onset of some numerical instabilities for the curves labeled
``IR'', in particular at small and large 
angles. This limits at present the numerical application of the
integral representation method.  

\begin{figure}[ht]
\includegraphics[width=.7\columnwidth]{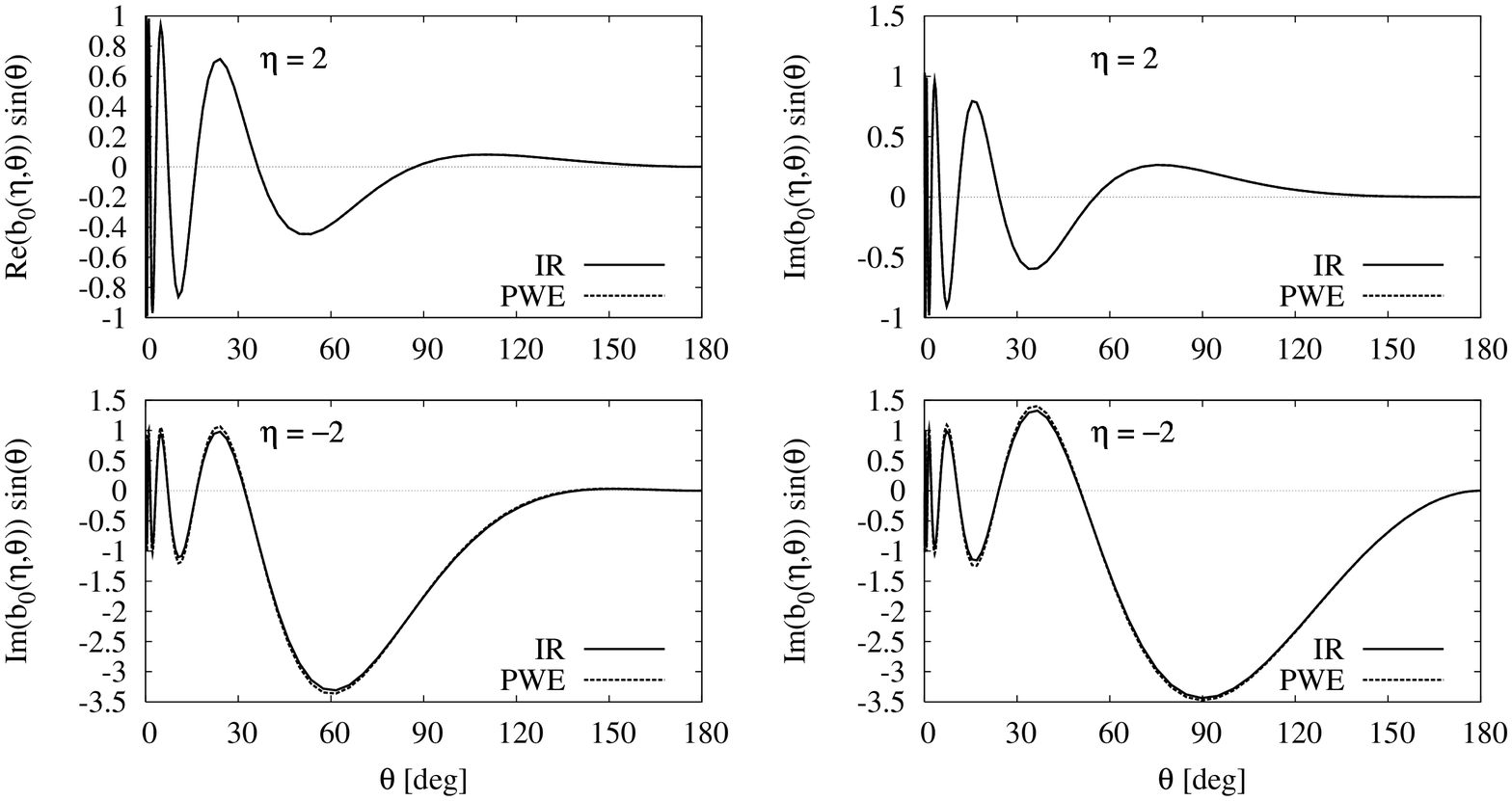}
\caption{Spin-orbit amplitude $b_{0}(\eta,\theta)$ 
for $\eta=\pm2$ for the integral
representation (IR) and the partial wave expansion (PWE).} 
\label{Fig_comparison_b0LS} 
\end{figure}

A comparison of the two methods for the spin-orbit amplitude is shown
in Fig.~\ref{Fig_comparison_b0LS}, again for $\eta= \pm 2$, where one
readily notes very good agreement. The same numerical problem of the
integral representation method arises also in this case for large
negative $\eta$.

%%%%%%%%%%%% appendix %%%%%%%%%%%%%%%%%%%%%%%%%%%%%%

\end{document}